\documentclass[a4paper,11pt]{article}
\usepackage[utf8]{inputenc}
\usepackage[english]{babel}

\usepackage{jheppub}

\usepackage{graphicx}
\usepackage{xcolor,color,overpic}
\usepackage{caption}
\usepackage{subcaption}

\usepackage{siunitx}
\let\svqty\qty
\usepackage{physics}
\let\qty\svqty
\usepackage{amsmath,amssymb,amsfonts,mathrsfs}
\def\be#1\ee{\begin{align}#1\end{align}} 

\usepackage{hyperref}
\hypersetup{colorlinks=true,linkcolor=red,citecolor=blue,urlcolor=black,pdfauthor={Jacopo Mazza}}

\usepackage[capitalize]{cleveref}
\crefname{equation}{eq.}{eqs.}
\crefname{figure}{figure}{figures}
\crefname{table}{table}{tables}
\crefname{subequation}{eqs.}{eqs.}
\labelcrefformat{subequation}{#2(#1)#3}
\crefformat{subequations}{#2(#1)#3}
\crefname{section}{section}{sections}
\crefname{appendix}{appendix}{appendices}

\newcommand{\SISSA}{\affiliation{SISSA, International School for Advanced Studies, \\
via Bonomea 265, 34136 Trieste, Italy}}
\newcommand{\InfnTS}{\affiliation{INFN, Sezione di Trieste,\\
via Valerio 2, 34127 Trieste, Italy}}
\newcommand{\IFPU}{\affiliation{IFPU, Institute for Fundamental Physics of the Universe,\\
via Beirut 2, 34014 Trieste, Italy}}

\usepackage{comment}
\usepackage{soul} 

\newcommand{\aet}{{\ae}ther}
\newcommand{\aem}{\text{\ae}}

\newcommand{\uh}{\text{UH}}
\newcommand{\kh}{\text{KH}}

\title{Regular Black Holes and Horizonless Ultra-Compact Objects in Lorentz-Violating Gravity}

\author{Jacopo Mazza}
\author{and Stefano Liberati}

\SISSA\IFPU\InfnTS

\emailAdd{jmazza@sissa.it}
\emailAdd{liberati@sissa.it}

\abstract{There is growing evidence that Ho\v{r}ava gravity may be a viable quantum theory of gravity.
It is thus legitimate to expect that gravitational collapse in the full, non-projectable version of the theory should result in geometries that are free of spacetime singularities. 
Previous analyses have shown that such geometries must belong to one of the following classes: simply connected regular black holes with inner horizons; non-connected black holes ``hiding'' a wormhole mouth (black bounces); simply connected or non-connected horizonless compact objects.
Here, we consider a singular black hole in the low-energy limit of non-projectable Ho\v{r}ava gravity, i.e.~khronometric theory, and describe examples of its possible ``regularisations'', covering all of the viable classes.
To our knowledge, these examples constitute the first instances of black holes with inner universal horizons, of black bounces and of stars with a de Sitter core in the context of Lorentz-violating theories of gravity.
}


\begin{document}

\maketitle \flushbottom

\section{Introduction}\label{sec:Introduction}

General relativity (GR) is notoriously non-renormalisable at the perturbative level \cite{t_hooft_one_1974,goroff_ultraviolet_1986}. 
Though this is not the only reason why a quantum theory of gravity remains elusive, it still represents an important technical impediment. 
To circumvent the issue, Ho\v{r}ava \cite{horava_quantum_2009} proposed a field theory of gravity in which power-counting renormalisability is manifest thanks to the addition, to the action of gravity, of terms that are higher-order in \emph{spatial} derivatives. 
This choice improves the ultra-violet (UV) behaviour of propagators while ensuring that no ghosts are introduced; but it comes at the cost of breaking diffeomorphism invariance and, locally, Lorentz invariance. 
In the following years, several improvements to the original idea have been proposed, including a ``healthy'' extension known as non-projectable Ho\v{r}ava gravity \cite{blas_extra_2009,blas_healthy_2010, blas_models_2011}, in which diffeomorphism invariance is restored via the introduction of an irrotational, unit-norm, timelike vector field --- the so-called \aet. 
There are now strong indications --- \cite{bellorin_brst_2022,bellorin_cancellation_2022,barvinsky_beta_2022,barvinsky_towards_2019,barvinsky_hovrava_2017,barvinsky_renormalization_2016} and references therein --- that this version of the theory is perturbatively renormalisable: if this property is confirmed, non-projectable Ho\v{r}ava gravity will represent an example of a consistent quantum theory of gravity in four spacetime dimensions.

At low energies, one can neglect all but the lowest-dimensional operators. 
The theory thereby obtained has an action of the form
\be\label{eq:action}
S = - \frac{1}{16\pi G}\int \dd[4]{x} \sqrt{-g} \left[R + \lambda (\nabla_a u^a)^2 + \beta \nabla_a u^b \nabla_b u^a + \alpha a^a a_a\right]\, ,
\ee
where $a_a = u^b\nabla_b u_a$ is the \aet's acceleration and $\alpha, \ \beta, \ \lambda$ are three dimensionless couplings.
This action coincides with that of Einstein--\aet~theory \cite{jacobson_gravity_2001, jacobson_einstein-aether_2008}, when the \aet~is constrained to be hypersurface-orthogonal \cite{jacobson_extended_2010,jacobson_undoing_2014} --- this justifies the choice of terminology for $u_a$.
However, since hypersurface orthogonality restricts the number of physical degrees of freedom, the theory specified by \cref{eq:action} goes by its own name: khronometric theory.

The parameters $\alpha, \ \beta, \ \lambda$ are tightly constrained by observations \cite{franchini_relation_2021}: $\abs{\beta} \lesssim 10^{-15}$ and either $\abs{\alpha} \lesssim 10^{-7}$ with $\lambda$ unconstrained or $\abs{\alpha}\lesssim \num{0.25d-4}$ with $\lambda \approx \alpha/(1-2\alpha)$. 
Moreover, $\lambda>0$ to avoid ghosts. 
Since $\alpha$ and $\beta$ seem both very small, one may at times consider a ``minimal theory'' in which they are set to zero exactly, while $\lambda$ remains free.

Due to hypersurface orthogonality, the \aet~can be expressed as
\be \label{eq:ugradT}
u_a = \frac{\nabla_a T}{N} \qq{with} N = \sqrt{\nabla_b T\nabla^b T}\, ,
\ee
and $T$ a scalar field called khronon. 
The khronon's level sets are three-dimensional, everywhere-spacelike hypersurfaces that provide a time foliation with a preferred status.
The existence of a preferred foliation --- or equivalently of a preferred reference frame, in this case the one provided by the \aet~--- is a direct consequence of the Lorentz-violating nature of Ho\v{r}ava gravity, and it opens the door to the existence of modified dispersion relations.
Indeed, khronometric theory allows for superluminal propagation of signals and even contains an instantaneous mode that travels at infinite speed.

In such a context, it would be natural to expect that the notion of black hole had no meaning.
Surprisingly, the theory does admit black hole solutions \cite{eling_black_2010, barausse_black_2011, blas_horava_2011, barausse_slowly_2013, barausse_no-go_2012, barausse_slowly_2016, oost_spherically_2021, bhattacharjee_gravitational_2018, berglund_mechanics_2012,janiszewski_charged_2014}. 
The moral equivalent of the event horizon is the so-called {universal} horizon (UH): a compact constant-khronon surface that traps modes of any speed --- even the instantaneous one. 
When the spacetime is stationary, meaning that there exists a Killing vector field $\chi^a$ that is timelike at infinity, the UH is characterised by the conditions \cite{bhattacharyya_causality_2016}
\be\label{eq:uh}
\eval{u_a \, \chi^a}_{\uh} = 0 \qq{and} \eval{a_a \, \chi^a}_{\uh} \neq 0\, .
\ee
UHs are thus different, in general, from the more familiar Killing horizons (KHs) --- defined as null hypersurfaces on which $\chi^a \chi_a =0$. Note that, although KHs are not causal horizons and play no particular role in theories with broken Lorentz invariance, \cref{eq:uh} implies that if a UH exists a KH must too: since $u_a$ is always timelike and $\chi^a$ is timelike at infinity, $u_a \chi^a = 0$ entails that $\chi^a$ must become null somewhere.

Known black hole solutions harbour a spacetime singularity at their centre, exactly as their general-relativistic counterparts do. 
One can conjecture, however, that these singularities might be ``cured'' by properly taking into account all the higher-dimensional operators that define the full theory. 
In this scenario, the end state of gravitational collapse would be more appropriately described by a non-singular (or regular) configuration, consisting of a non-singular metric {and a non-singular \aet~flow}. 
Solutions of this kind are however still lacking, and their derivation is probably going to be challenging. 

On the other hand, a lot is known about singularities (and how to avoid their formation) in the context of purely metric theories of gravity: it is thus natural to wonder whether this know-how could guide the search for non-singular configurations in Ho\v{r}ava gravity. 
In particular, if one assumes that pseudo-Riemannian geometry provides a good description of the spacetimes resulting from quantum gravitational regularisation at late times, Penrose's singularity theorem \cite{penrose_gravitational_1965} can be used to compile a classification of all the non-singular geometries that gravitational collapse could result in \cite{carballo-rubio_geodesically_2020, carballo-rubio_opening_2020}. 
This approach is purely geometric and therefore largely theory-agnostic. 

Surprisingly, the resulting list of viable spacetimes is remarkably short. 
In essence, once a trapping horizon is formed somewhere in the spacetime, there are only two options for avoiding  an inner singularity: either the geometry is endowed with an ``inner'' trapping horizon; or the singularity is replaced by a wormhole mouth, which is ``hidden'' behind the trapping horizon. 
Note that, while in the former case the spacetime is simply connected, the latter case is characterised by the existence of a minimum radius that renders the spacetime topologically different.
For this reason, we will refer to these alternatives as \emph{connected} and \emph{non-connected} regularisation, respectively. 
The only way to avoid either of them is to prevent the formation of the trapping horizon altogether. In principle, one can still consider both connected and non-connected configurations without horizons: while a connected horizonless object is a star (though possibly of an exotic kind), a non-connected one is a traversable wormhole.

The analysis of \cite{carballo-rubio_geodesically_2020} was performed under the assumption of Lorentz invariance but it can be extended to the framework of Lorentz-breaking theories \cite{carballo-rubio_geodesically_2022}. 
Despite some technical subtleties, the main result carries over: non-singular configurations are either simply connected or non-connected; and they might display a universal trapping horizon (the moral equivalent of the trapping horizon) or not --- but if they are connected and with horizon, a second universal inner horizon must exist.

Such classification is best suited for describing dynamical situations, namely the collapse of some energy density leading to the formation of a compact object. 
In particular, \cite{carballo-rubio_geodesically_2020,carballo-rubio_geodesically_2022} assume global hyperbolicity and therefore do not admit, for instance, stationary simply connected regular black holes. 
Yet, if the evolution of the geometry is characterised by a timescale that is much longer than any other relevant timescale, a stationary non-singular spacetime might provide an approximate description that is sufficiently accurate for phenomenological purposes.

This line of reasoning has given rise to a thriving industry, fuelled by recent observational achievements, aimed at constructing models of non-singular geometries. 
Models of simply connected regular black holes have a longer history, dating back to the work of Bardeen in 1968 \cite{bardeen_non-singular_1968}, and are therefore more abundant in the literature  --- see e.g.~\cite{ansoldi_spherical_2008, maeda_quest_2022, sebastiani_remarks_2022} end references therein. 
Several (though fewer) instances of wormholes whose mouths are hidden behind an horizon exist as well \cite{simpson_black-bounce_2019, simpson_vaidya_2019,lobo_novel_2021,franzin_charged_2021, bronnikov_regular_2006, bronnikov_regular_2007,bronnikov_field_2022}. 

Only very recently \cite{carballo-rubio_connection_2022}, it was realised that the same models can describe horizonless objects too: typically, horizons only exist when the parameters that specify the model fall within a given range; but when they do not, the corresponding geometries are still perfectly viable. 
In particular, the connected regular black holes are counterparts of compact stars that usually have a de Sitter core and are therefore instances of gravastar-like objects \cite{mazur_gravitational_2004,mottola_effective_2022}.

These models, which are usually constructed \textit{ad hoc} and without referring to physically well-motivated theories, implicitly assume that all the relevant information concerning the geometry is encoded in the metric. 
In theories like Ho\v{r}ava gravity, in contrast, the preferred foliation has a crucial, genuinely physical role that is not entirely played by the metric. 

The goal of this paper, therefore, is to construct explicit examples of non-singular geometries, connected and non-connected, in the context of low-energy Ho\v{r}ava gravity. 
Such examples will be ``regularisations'' of a known singular solution that represents a black hole in the phenomenologically viable branch of the theory; they will consist of a metric and an \aet~flow, both of which will be free of singularities. 
They will not constitute solutions of khronometric theory in vacuum, nor with any simple form matter; yet, they will display all the key features that are expected for exact non-singular solutions of both khronometric theory and the full Ho\v{r}ava gravity. 
In particular, these configurations will exhibit pairs of inner/outer UHs, hidden wormholes and (gravastar-like) compact stars with de Sitter core --- all features that, to our knowledge, have never been described before in this context.
We hope our work will inform the search for regular black hole solutions in a promising candidate theory of quantum gravity; and spur further investigations into the phenomenology of gravity theories with broken Lorentz invariance.

Non-singular black holes have been searched for, but not found, in $(2+1)$ projectable Ho\v{r}ava gravity in \cite{lara_black_2021}. 
Spherical stars in Einstein-\aet~and khronometric theory have been investigated in \cite{eling_spherical_2010}, while examples of wormholes are given in \cite{chojnacki_finite_2021, lin_ellis_2022}; however, the \aet~flow considered in these references is often assumed to be parallel to the Killing vector: our analysis is therefore substantially different.

\bigskip

The paper is structured as follows. 
\Cref{sec:theo} is an introduction to the singular solution we take as a starting point for constructing our non-singular geometries. 
Such geometries are introduced in 
\cref{sec:regularise} and characterised in the following sections.
In particular, \cref{sec:hor} investigates horizons (Killing and universal);
\cref{sec:causal} describes the causal structure; 
\cref{sec:matter} quantifies the deviations away from the vacuum of the khronometric theory. 
Finally, \cref{sec:conc} reports our conclusions.
We also include \cref{app:scalars,app:expansions}, which provide further details on the non-singular configurations.

\section{A (singular) black hole solution}\label{sec:theo}

The equations of motion obtained by varying \cref{eq:action} with respect to $\var g_{ab}$ and $\var T$ can be written as:
\be 
\mathcal{G}_{ab}  = 0 \, , \label{eq:eomg}\\
\nabla_a \left( \frac{\mathcal{A}^a}{N} \right) = 0\, .\label{eq:eomAE}
\ee
\Cref{eq:eomg} is the equivalent of the Einstein's equation, indeed one can write $\mathcal{G}_{ab}= G_{ab} - T^\aem_{ab}$, with $G_{ab}$ the Einstein's tensor and  $T^\aem_{ab}$ the stress-energy tensor (SET) of the \aet. 
\Cref{eq:eomAE} is the equation of motion of the khronon: the vector $\mathcal{A}^a$ is built out of $u_a$ and its derivatives and is orthogonal to the \aet~$u_a \mathcal{A}^a =0 $.
To include matter, one can add the matter action $S_\text{mat}$ to \cref{eq:action}. 
This yields source terms that appear on the right-hand side of \cref{eq:eomg,eq:eomAE}.

In this paper, we will investigate spherically symmetric and static spacetimes, and assume that the same symmetries extend to the \aet~field. 
This implies the existence of a Killing vector $\chi^a$, timelike at spatial infinity, along which both the metric and \aet~are Lie-dragged.
Adopting in-going Eddington--Finkelstein coordinates, we can generically write the metric and the \aet~as

\be
ds^2 &= F(r) \dd{v}^2 - 2 \dd{v}\dd{r} - R^2(r)\dd{\Omega}^2\, , \label{eq:met}\\
u^a\partial_a &= A(r) \partial_v + y(r) \partial_r\, . \label{eq:aet}
\ee

These equations define the variables that we will be using for the remainder of the paper: $F(r)$ and $R(r)$ parametrise the metric degrees of freedom, while $A(r)$ and $y(r)$ are the two non-zero components of the \aet. 

Note that, since the \aet~has unit norm, the following relation holds:
\be 
y = - \frac{1-A^2F}{2A}\, .
\ee
With these notations, the projection of the \aet~along the Killing vector is
\be 
u_a\,\chi^a =  \frac{1+A^2F}{2A}\, .
\ee

An exact solution is known for $\alpha = 0$ \cite{berglund_mechanics_2012}.\footnote{Another exact solution can be found for $\beta+\lambda =0$.} In our coordinates, it is given by\footnote{Reference \cite{berglund_mechanics_2012} only reports the plus sign, although the equations of motion actually allow for both. However, the square root, with its sign, coincides with $u_a \, \chi^a$, which must become negative upon crossing the universal horizon \cite{del_porro_time_2022}. Hence, the choice of \cite{berglund_mechanics_2012} is in fact ill-behaved at the universal horizon; all their conclusions still stand, despite this clarification.}  
\be 
F &= 1-\frac{r_0}{r} - \beta \frac{r_{\aem}^4}{r^4}\, , \qquad R(r) = r\, , \label{eq:exact_met}\\
A(r) &= \frac{1}{F(r)} \left[-\frac{r_\aem^2}{r^2} \pm \sqrt{F(r) + \frac{r_\aem^4}{r^4}} \right]\, , \label{eq:exact_A}
\ee
where $r_0$ is twice the Arnowitt--Deser--Misner (ADM) mass of the spacetime (measured in appropirate units) and $r_\aem$ is another, \textit{a priori} independent, integration constant. 
Depending on the relative magnitude of $r_0$ and $r_\aem$, the quantity
\be 
F(r) + \frac{r_\aem^4}{r^4}
\ee
may become negative, thus rendering $A(r)$ ill-defined. 
One can further check that this quantity coincides with $(u_a \, \chi^a)^2$, hence its zeroes correspond to UHs. 
Only when the parameters satisfy
\be \label{eq:fine_tune}
r_\aem = \frac{r_0}{4} \left( \frac{27}{1-\beta} \right) ^{1/4} 
\ee 
does the solution describe a black hole with one universal horizon located at
\be 
r_\uh = \frac{3}{4}r_0\, ;
\ee
when $r_\aem$ is larger than this value no UHs exist, when instead it is smaller the \aet~flow is ill-defined.
Assuming this fine-tuned choice, as we will do for the rest of the paper, one can write
\be \label{eq:exact_uchi}
u_a \, \chi^a = \frac{1}{r^2} \left( r - \frac{3}{4}r_0 \right) \sqrt{r^2 + \frac{r_0}{2} r + \frac{3r_0^2}{16} }\, .
\ee 
The signs have been chosen so that this quantity tends to one at spatial infinity but changes sign upon crossing the universal horizon --- as it must \cite{del_porro_time_2022}. 
This corresponds to choosing the plus sign in \cref{eq:exact_A} outside of the UH and the minus inside.

The UH has an associated surface gravity that sets the temperature of the analogue of Hawking's radiation, in a way similar to the surface gravity of horizons in GR (see e.g.~\cite{Herrero-Valea:2020fqa,del_porro_time_2022,del_porro_gravitational_2022} and references therein). 
It is defined as
\be\label{eq:kappa}
\kappa_\uh = - \frac{1}{2} a_a \, \chi^a \, ,
\ee
which on the singular solution evaluates to
\be\label{eq:kappa:es}
\kappa_\uh^\text{sing.} = \frac{2\sqrt{2}}{3 \sqrt{3} r_0 \sqrt{1-\beta}}\, .
\ee

The metric also exhibits a Killing horizon (KH), associated with the zero of $F(r)$. 
Clearly, when $\beta=0$ the metric reduced to that of Schwarzschild and the KH is located at $r_\kh=r_0$. 
More generally, one can write $F(r)=0$ as
\be\label{eq:horbeta}
r^3(r-r_0) - \left[\frac{27}{256} \frac{\beta}{1-\beta} r_0^4 \right]=0 \, ;
\ee
hence, one can deduce that the KH moves towards larger values of $r$ as $\beta$ increases (we always assume $\beta < 1$), i.e. $r_\kh \geq r_0$. 
Thus, the KH always encloses the UH. The equation does not have any more roots. 
Note that $A(r)$ is well-behaved at the KH, as can be verified by expanding close to $r=r_\kh$:
\be
A (r) &= \frac{1}{F'\left(r_\kh \right) \left( r-r_\kh \right) } \left[ \frac{r_\kh^2 F'\left(r_\kh\right)}{2 r_\aem ^2} \left(r-r_\kh \right) \right] + \order{\left( r-r_\kh \right)^2}\\
 &=   \frac{r_\kh^2}{2 r_\aem^2} + \order{\left( r-r_\kh \right)^2}\, .
\ee

This metric is singular at $r=0$, as one can check e.g.~by evaluating the Kretschmann scalar: 
\be
R_{abcd} R^{abcd} = 12 \frac{r_0^2 r^6 + 10\beta r_0 r_\aem ^4 r^3 + 39 \beta^2 r_\aem ^8}{r^{12}} \, .
\ee
The components of the \aet~also seem ill-defined at that point, although this statement relies on the choice of coordinates. 
To check that the \aet~flow is in fact singular at $r=0$ one should characterise it in terms of scalar quantities. 
Since the \aet~constitutes a timelike non-geodesic congruence, a rather natural choice is to describe it in terms of its optical scalars: \footnote{The term ``optical scalars'' is usually reserved for null geodesic congruences. We are abusing this terminology, hopefully without confusion.} the expansion, the square of the symmetric shear and  the square of the antisymmetric twist. 
Further details can be found in \cref{app:scalars}.

\section{Regularisations of the singularity}\label{sec:regularise}

As mentioned in \cref{sec:Introduction}, there are only two qualitatively different ways of avoiding, i.e.~``regularising'', the central singularity: we have referred to these alternatives as connected and non-connected regularisation.
The connected regularisation corresponds to a physical scenario in which gravity effectively becomes weaker and ``turns off'' at $r=0$. 
Metrics that exhibit such behaviour can be built by modifying a singular solution (typically Schwarzschild) in a simple way.
The non-connected regularisation instead does not correspond to a weakening of gravity. 
On the contrary, gravity becomes so strong that it induces a change in the topology of spacetime. 
As a consequence, a finite region containing $r=0$ gets excised from the spacetime: this alternative is therefore characterised by the existence of a minimal length scale corresponding, roughly speaking, to the radius of the smallest sphere centred at $r=0$. 
Despite the different topology, metrics portraying this scenario can be built by modifying a singular solution too.

To be explicit and as clear as possible, in the following we will explore two specific examples, one connected and one not. 
Most of the qualitative considerations, however, hold true in general.

\Cref{app:expansions} reports further details, using a characterisation in terms of two-dimensional congreunces in order to make contact with the language of \cite{carballo-rubio_geodesically_2022}.

\subsection{Connected regularisation}

A simple way to construct a simply connected non-singular metric, starting from a singular one, is to upgrade the parameter $r_0$ to a function of the radius $r_0(r)$. 
Remarkably, when this replacement is performed on \cref{eq:exact_met} and \cref{eq:exact_A}, the resulting \aet~becomes regular too. 
Explicitly, the metric components of \cref{eq:met} and the \aet~one of \cref{eq:aet} become
\be \label{eq:regConm}
F(r) &= 1-\frac{r_0(r)}{r} - \beta \frac{r_\aem ^4(r)}{r^4}\, , \qquad R(r) = r\, ,\\ \label{eq:regConae}
A(r) &= \frac{1}{F(r)} \left[-\frac{r_\aem ^2(r)}{r^2} + \frac{1}{r^2} \left( r - \frac{3}{4}r_0(r) \right) \sqrt{r^2 + \frac{r_0 (r) }{2}r + \frac{3r_0^2(r)}{16} } \right]\, ,\\ 
r_\aem (r) &= \frac{r_0 (r)}{4} \left( \frac{27}{1-\beta} \right)^{1/4} \, .
\ee

The function $r_0(r)$ is arbitrary, except for a minimum set of requirements \cite{ansoldi_spherical_2008, maeda_quest_2022, sebastiani_remarks_2022, frolov_notes_2016}: it should be ``well-behaved'', in the sense that it should not introduce new singularities (this is guaranteed if e.g.~$r_0(r)>0$); 
it should not spoil asymptotic flatness, i.e.~$\lim_{r\to\infty} r_0(r) = 2M$ with $M$ the ADM mass; 
and, crucially, it must be $r_0(r) = \order{r^3}$ close to $r=0$. 
This last requirement makes the components of the metric manifestly regular at the origin and prevents divergences in any scalar polynomial built out of the Riemann tensor and the metric. 
Expanding close to $r=0$, one has
\be 
F(r) = 1 - c r^2 + \order{r^3}\, ,
\ee
showing that the geometry of the inner core is asymptotically de Sitter (anti-de Sitter) if $c>0$ ($c<0$) or Minkowski if $c=0$.

The components of the \aet~are now manifestly regular, too. 
In particular,
\be 
A(r) = 1 + \frac{c}{2}r^2 + \order{r^3} \qq{and} y(r) = \order{r^3}\, ,
\ee
i.e.~in the limit $r\to0$ the \aet~coincides with the Killing vector, up to corrections of order $\order{r^2}$. 
This is precisely the trivial \aet~flow that one would expect in a maximally symmetric space.
The first derivatives of $F(r)$ and $A(r)$ are similarly well-behaved close to $r=0$, which ensures that all the optical scalars characterising the \aet~congruence are regular too --- details can be found in \cref{app:scalars}.

Note, incidentally, that the geometry described by \cref{eq:regConm} is certainly non-singular, in general, only for $r\geq0$. 
If one allows the coordinate $r$ to become negative, one might still encounter spacetime singularities \cite{zhou_geodesic_2022,lan_singularities_2022} --- in the form of divergences in the curvatures or in the sense of geodesic incompleteness. 
In order to interpret \cref{eq:regConm} as a non-singular black hole, therefore, one must limit the domain of $r$ to $[0, +\infty)$. 
Clearly, this is coherent with the interpretation of $r$ as a radius, and with the fact that $r=0$ is, at any given $v$, a point (i.e.~a degenerate, zero-radius sphere).

Many explicit forms of $r_0(r)$ have been proposed and the corresponding metrics have been extensively studied: all of them are characterised by at least one additional parameter, usually carrying the dimensions of a length, upon which $r_0(r)$ depends continuously. 
Often, $r_0(r)$ reduces to a constant for some particular values of the parameters (typically zero). 
In this limit, the regularisation is undone.

In what follows, we will present calculations for a particular choice of $r_0(r)$ introduced by Hayward \cite{hayward_formation_2006},
\be \label{eq:Hayward}
r_0(r) = 2M \frac{r^3}{r^3 + 2M \ell^2}
\ee
(we have called $\ell$ the additional parameter; note that $r_0=2M$ for $\ell=0$), but the features we will describe are generic: other well-studied examples, e.g.~the Bardeen \cite{bardeen_non-singular_1968} or Dymnikova \cite{dymnikova_vacuum_1992} metrics, yield very similar results.

\subsection{Non-connected regularisation}

Many instances of wormhole exist in the literature (see e.g.~\cite{visser_lorentzian_1996}). 
In the past, the attention has mostly focused on ``traversable'' wormholes, i.e.~wormholes with a timelike mouth that can be traversed in both directions. 
However, there has been recently a growing interest towards ``hidden'' wormholes, i.e.~wormholes whose mouths are shielded by trapping horizons. 
In the presence of a single outer horizon such mouth is not traversable, being spacelike, and indeed the metric can be more precisely characterised as a ``black-bounce" being endowed with a minimum radius.

The example that we investigate here is a very simple black-bounce geometry proposed by Simpson and Visser \cite{simpson_black-bounce_2019}.
The original metric is a slight modification of the Schwarzschild one, formally obtained by replacing any instance of $r$ with $\sqrt{r^2 + \ell^2}$.
When this trick is applied to the singular solution \cref{eq:exact_met,eq:exact_A}, the metric components of  \cref{eq:met} and the \aet~one of \cref{eq:aet} take the form
\be
F(r) &= 1-\frac{r_0}{\sqrt{r^2 + \ell^2}} - \beta \frac{r_\aem ^4}{(r^2 + \ell^2)^2}\, , \qquad R(r) = \sqrt{r^2 + \ell^2}\, , \label{eq:SVmet}\\
A(r) &= \frac{1}{F(r)} \left[-\frac{r_\aem ^2}{r^2 + \ell^2} + \frac{1}{r^2 + \ell^2} \left( \sqrt{r^2 + \ell^2} - \frac{3}{4}r_0 \right) \sqrt{(r^2 + \ell^2) + \frac{r_0 \sqrt{r^2 + \ell^2}}{2} + \frac{3r_0^2}{16} } \right] \, . \label{eq:SVaet}
\ee
We are still assuming $r_\aem =  27^{1/4} (1-\beta)^{-1/4} (r_0/4)$, without $r$-dependence, as in the singular solution.

In this example, regularity is manifest, since all components of both the metric and the \aet~approach a finite non-zero limit as $r\to0$.
(Details on the \aet's optical scalars can be found in \cref{app:scalars}.)
Notably, $R(r) = \ell + \order{r^2}$, meaning that, at any given time $v$, $r=0$ is not a point; rather, it is a sphere of surface area $4\pi \ell^2$.
It is therefore clear how this example could be generalised:  any other $R(r)$ that attains a finite value in $r=0$ works as fine --- provided one writes $F\left( R(r) \right)$ and $A \left( R(r) \right)$.

Since the metric and the \aet~are invariant under $r\to-r$, one can extend the domain of the coordinate $r$ to $(-\infty, +\infty)$. 
Hence, the geometry should be interpreted as representing a wormhole whose asymptotically flat ends are at $r\to+\infty$ and $r\to-\infty$ and whose mouth, a sphere of surface area $4\pi\ell^2$, is located at $r=0$. 
We will often call ``our universe'' the $(r>0)$-patch and ``other universe'' the $(r<0)$-patch.

It is useful to express \cref{eq:SVmet,eq:SVaet} in terms of a new coordinate $\varrho = \sqrt{r^2 + \ell^2}$. We have
\be
ds^2 &= F(\varrho) \dd{v}^2 - 2 \delta(\varrho) \dd{v}\dd{\varrho} - R^2(\varrho) \dd{\Omega}^2\, ,\\
u^a \partial_a &= A(\varrho) \partial_v + \frac{y(\varrho)}{\delta(\varrho) }\partial_\varrho\, , 
\ee
where 
\be
\delta(\varrho) = \dv{r}{\varrho} = \frac{\varrho}{\sqrt{\varrho^2-\ell^2}}
\ee
and 
\be
F(\varrho) &= 1 - \frac{r_0}{\varrho} - \beta \frac{r_\aem^4}{\varrho^4}\, , \quad R(\varrho) = \varrho\, ,\\
A(\varrho) &= \frac{1}{F(\varrho)} \left[ -\frac{r_\aem^2}{\varrho^2} + \frac{1}{\varrho^2} \left(\varrho-\frac{3}{4}r_0\right)\sqrt{\varrho^2+\frac{r_0 }{2} \varrho + \frac{3}{16}r_0^2}\right]\, .
\ee
The new coordinate $\varrho$ has a simple physical interpretation: it is the aerial radius, i.e.~surfaces of constant $\varrho$ (and $v$) are spheres with area $4\pi \varrho^2$. 
In this coordinate, the components of the metric and of the \aet~look identical to those of the singular case, except for $\delta$; however, $\varrho$ can never reach zero, since $\min(\varrho) = \varrho\left( r=0 \right)=\ell>0$, and the geometry thus remains free of spacetime singularities. 
There is now a coordinate singularity at the throat $\varrho=\ell$, hence this coordinate system can only describe one of the two universes at a time.

\section{Horizons}\label{sec:hor}

The regularised metrics introduced above, similarly to the singular solution, exhibit KHs located at the solutions of $F(r) = 0$.
These horizons are still surfaces of infinite redshift/blueshift for matter that is minimally coupled to the metric and uncoupled to the \aet. 
However, because of the breaking of local Lorentz invariance, they are not causal horizons since the presence of superluminal signals affects the causal structure \cite{carballo-rubio_causal_2020}. 
As mentioned above, in khronometric gravity the role of causal horizons is instead played by UHs (see e.g.~the relative discussion in~\cite{carballo-rubio_causal_2020} and references therein).

Nonetheless, it is easy to see that in the configurations we are exploring these structures are tightly related. 
Indeed, for both kinds of regularisation (as for the singular case) one can write
\be
\left(u_a \, \chi^a \right)^2 = F(r) + f(r) \qq{with} f(r) >0\, ,
\ee
so, since both $u_a \, \chi^a$ and $F$ are positive at infinity, $u_a \, \chi^a$ can reach zero only in a region in which $F(r)$ is negative. 
This means that UHs must necessarily lie in a ``trapped region'' --- as one would call it in GR. 
Hence, the presence of a UH implies the existence of a KH that encloses it. 
Moreover, $F(r)$ must be positive at $r=0$ in the connected case and at $r=-\infty$ in the disconnected case. 
This entails that KHs have to come in pairs and therefore, generically, UHs have too.

Thus, non-singular black hole geometries typically exhibit a nested structures of Killing and universal horizons. 
In the following subsections we will describe this structure in detail for the specific examples that we are exploring, but the above considerations can be proven to be general by exploiting the notion of the degree of a map. 
Moreover, in \cref{app:expansions} we present an alternative, local characterisation of UHs in terms of the expansions of two congruences, in a language that makes contact with \cite{carballo-rubio_geodesically_2022}.

\subsection{Connected regularisation}

We start by setting $\beta = 0$, for simplicity. KHs are given by $r = r_0(r)$ and UHs by $ 4r/3 = r_0(r)$.
Whether these equations admit solutions or not is a model-dependent question. 
When $r_0(r)$ is that of \cref{eq:Hayward}, for example, the answer depends on the value of the parameter $\ell$.
\begin{figure}
    \centering
    \includegraphics[width=.7\textwidth]{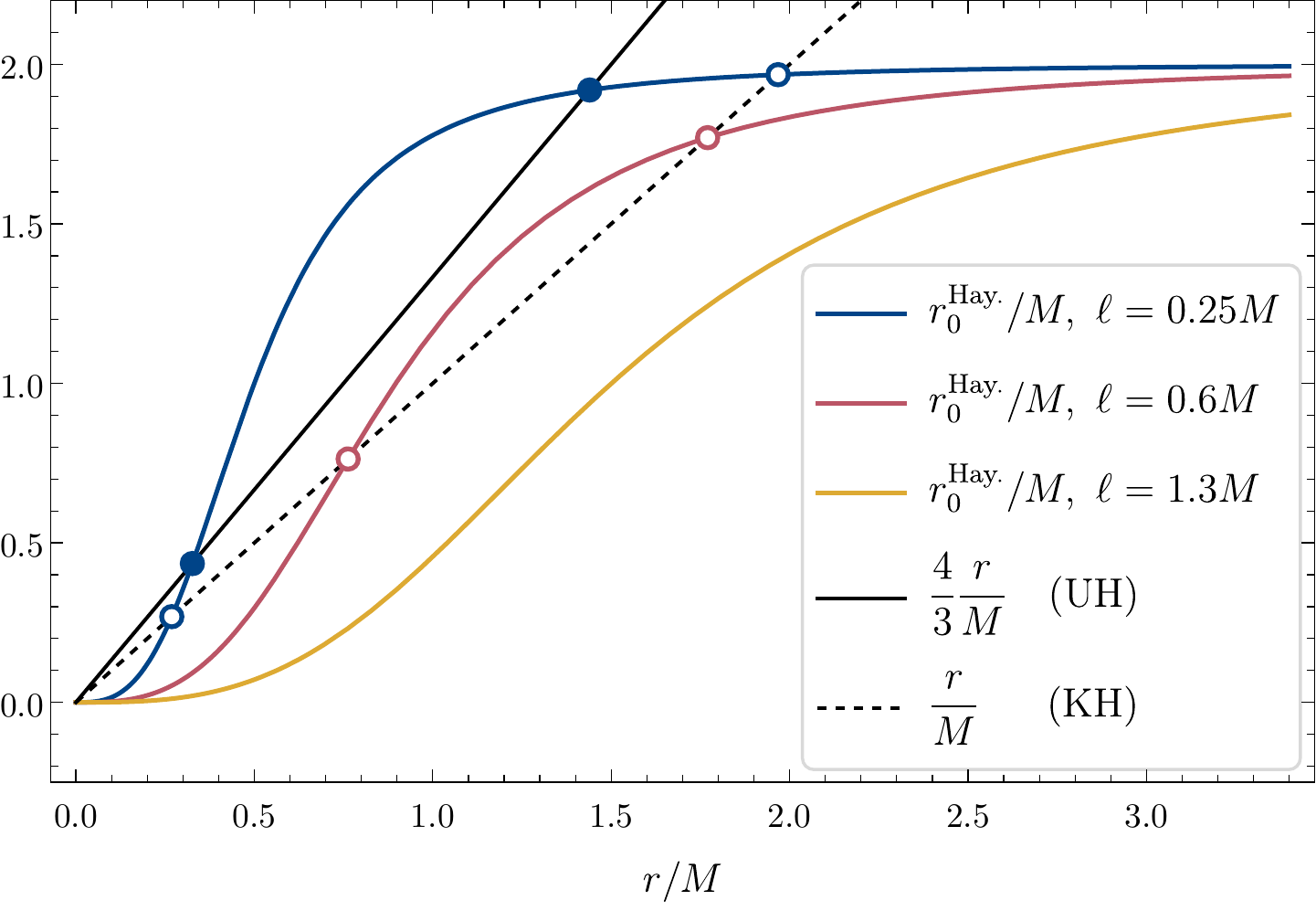}
    \caption{Plot of Hayward's choice of $r_0(r)$ for three values of the parameter $\ell$. Intersections with the straight (dashed) black line correspond to UHs (KHs) in the minimal theory $\alpha=\beta=0$.}
    \label{fig:hor}
\end{figure}

As an illustration, in \cref{fig:hor} we plot Hayward's $r_0(r)$ for three values of $\ell$; the plot also reports two straight lines, with slope equal to one (dashed line) and $4/3$ (solid line) respectively: the intersections of $r_0(r)$ with these lines determine the horizons. 
When $\ell$ is small, we can count two intersections with the dashed line and two with the solid one. 
Hence, this configuration presents two KHs and two UHs; coming from infinity, they are met in the following order: outer KH, outer UH, inner UH, inner KH.
As $\ell$ is increased, the curve relative to $r_0(r)$ moves towards the bottom-left corner of the picture: inner and outer horizons thus approach each other. 
They keep approaching until the two UHs merge into a single, degenerate UH; this happens at a threshold value of $\ell=M/2$ above which no UH exist. 
Similarly, the KHs keep approaching until they merge into a degenerate KH and then disappear: this second threshold corresponds to a higher value of $\ell = 4M/(3\sqrt{3})$.

Therefore, we can distinguish three qualitatively different regimes: a non-singular black hole regime, characterised by an inner/outer UH pair (as well as an inner/outer KH pair); an intermediate regime in which there are two KHs but no UHs; and a star-like regime with no horizons. 
Although technically a black hole is present only in the first regime, an object in the intermediate regime would still appear ``almost black'', given that low energy modes would linger for an extremely long time at the KH before being able to escape to infinity.

Reinstating the parameter $\beta$ does not greatly distort this picture, since its only effect is that of displacing the KHs. 
\Cref{eq:horbeta} remains valid upon replacing $r_0$ with $r_0(r)$, so increasing the value of $\beta$ shifts the outer KH outwards. 
The inner KH, instead, moves inwards. 
That is, increasing $\beta$ has the effect of pushing KHs further apart; this is the opposite effect one has by increasing the regularisation parameter $\ell$, which instead pushes KHs closer together. 
The location of UHs is unaffected by $\beta$.

In the black hole regime, the universal horizons each have a surface gravity.
Plugging \cref{eq:regConae} in the definition \cref{eq:kappa}, we get
\be
\kappa_\uh &= \eval{\frac{4-3r_0'(r)}{3 \sqrt{6} r_0 \sqrt{1-\beta}}}_{UH} \, ,
\ee
which should be evaluated at each of the UHs.
Note that when $r_0'=0$ we recover the result for the singular solution \cref{eq:kappa:es}. 

In the black hole and in the intermediate regime, the horizon radii provide an intuitive way of telling the ``size'' of the compact object. 
It would be useful to extend this notion to the star-like regime by defining an appropriate effective radius.
A particularly simple choice is to pick the unique $r_\star$ for which $F'(r_\star)=0$.
This is the radius of maximum (metric) redshift and thus quantifies the compactness of the star.
Moreover, in the limit in which $\ell$ approaches (from above) the threshold value for the KH's formation, $r_\star$ approaches the (degenerate) horizon radius.

Explicitly, we have
\be
F' = - \left( \frac{r_0}{r} \right)' \left[1 + 4 \frac{27}{256} \frac{\beta}{1 -\beta} \left( \frac{r_0}{r} \right) \right]\, ,
\ee
hence $F'=0$ reduces to
\be
r_0(r) - r r_0'(r) =0 \, ,
\ee
independently on $\beta$.
For Hayward's choice, we find
\be
r_\star = \left( 4M \ell^2 \right)^{1/3} \, .
\ee

\subsection{Non-connected regularisation}

As in the previous case, the existence and location of horizons is, strictly speaking, a model-dependent question. 
In the simple example that we are considering, the answer is determined by the only free parameter $\ell$. 
The discussion becomes particularly simple if one resorts to the coordinate $\varrho$.

KHs are solutions of 
\be
\varrho^3 (\varrho-r_0) - \left[\frac{27}{56} \frac{\beta}{1-\beta}r_0^4 \right] = 0\, ,
\ee
which is formally the same as \cref{eq:horbeta}. Call $\varrho_\kh$ the (unique) solution; in the $r$ coordinates, this corresponds to
\be
r_\kh^2 = \varrho_\kh ^2 - \ell^2\, . 
\ee
Hence, for $\ell < \varrho_\kh$ the spacetime has one KH per each universe, located at $r = \pm r_\kh$ with $r_\kh = \sqrt{\varrho_\kh^2-\ell^2}$; when instead $\ell > \varrho_\kh$ the spacetime has no KHs; the limiting case $\ell = \varrho_\kh$ corresponds to the two horizons coinciding with the wormhole mouth, which in this case is null. 
Similarly to the simply connected configuration, one can easily check that increasing $\ell$ makes the KH shrink, while increasing $\beta$ makes it larger.

For what concerns the UHs, instead, they are located at
\be
\varrho_\uh = \frac{3}{4}r_0\, .
\ee
That is, when $\ell < 3 r_0 /4$ there is one UH per each universe, located at $r=\pm r_\uh$ with $r_\uh = \sqrt{\varrho_\uh^2 - \ell^2}$; when instead $\ell > 3 r_0 /4$ there are no UHs. 
As before, the equality corresponds to a degenerate case for which the mouth of the wormhole coincides with the UH. 
Analogously to the previous case, increasing $\ell$ makes the UH shrink while $\beta$ has no effect at all; note that $r_\uh < r_\kh$.

The surface gravity of the UH has a particularly simple form:
\be
\kappa_\uh = \kappa_\uh^\text{sing.} \sqrt{1-\frac{\ell^2}{\varrho_\uh^2}}\, ,
\ee
where $\kappa_\uh^\text{sing.}$ is the surface gravity for the singular solution written in \cref{eq:kappa:es}. 
Thus, for $\ell = \varrho_\uh = 3r_0/4$ the UH is ``degenerate'' and the black hole is extremal, in the sense that its UH's surface gravity vanishes.

\section{Causal structure}\label{sec:causal}

In a Lorentz-violating theory of gravity like khronometric theory, the causal structure is not determined by null rays. 
Rather, the theory exhibits a preferred foliation, specified by constant-khronon surfaces: it is the embedding of the leafs of the foliation into the four-dimensional spacetime, therefore, that defines the causal structure. 

Due to spherical symmetry, the khronon does not depend on the angles in either of the spacetimes we constructed. 
Hence, we can visualise the causal structure by simply plotting, in an appropriate (time--radius) plane, the surfaces of constant khronon. 
The most natural definition of time is given in terms of the null coordinate $v$ as
\be 
\dd t_* = \dd v - \dd r\, ;
\ee
this is a (Killing-)``horizon-penetrating'' time.
The more familiar time $t$, given by $\dd t = \dd v - \dd r/F$, would not be appropriate, since the components of the metric and of the \aet~are singular at the KHs when expressed in terms of it.

In addition, we plot the flow of the \aet, written in its covariant form. 
Since the \aet~is by definition orthogonal to constant-khronon hypersurfaces, the information provided by these plots and by those representing constant-khronon surfaces is not independent but complementary. 
We report both, hoping this will benefit the reader.

\subsection{Connected regularisation}

\begin{figure}[htb!]
    \centering
    \subfloat[$\ell = 0.25M$.]{\includegraphics[width=0.45\textwidth]{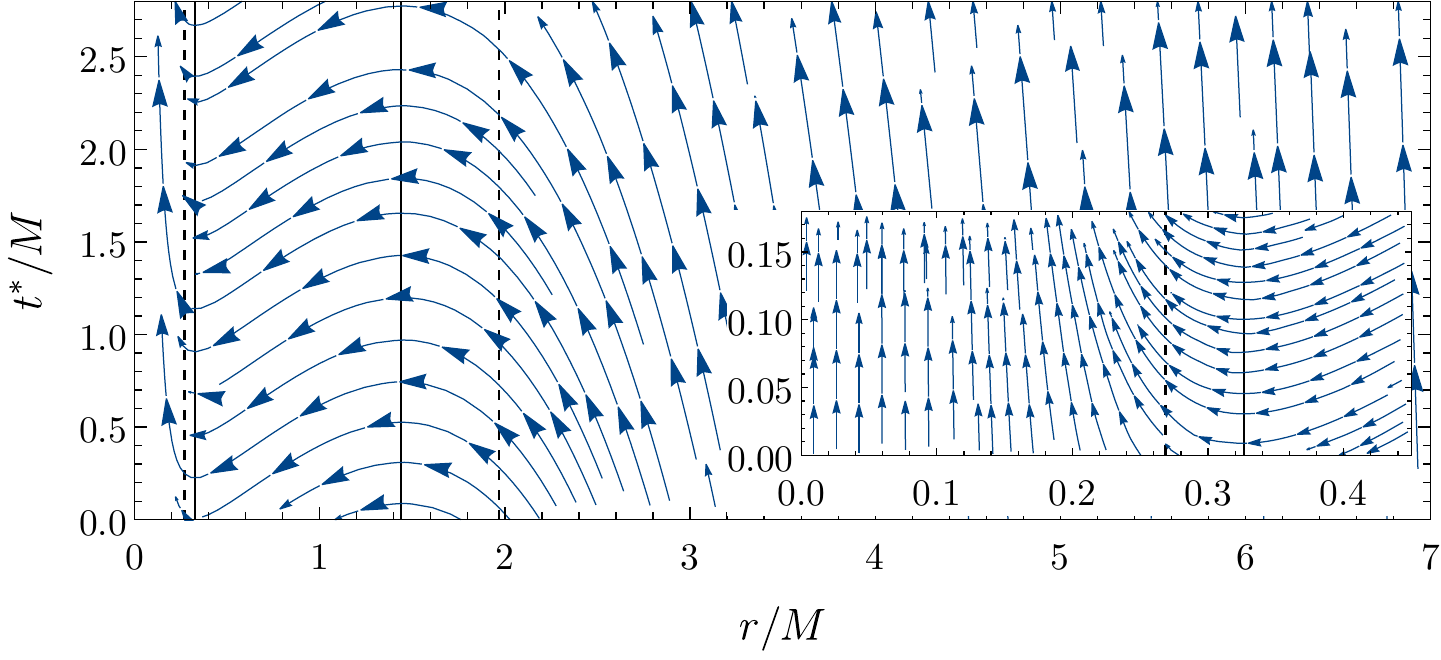}}
    \hfill
    \subfloat[$\ell = 0.25M$.]{\includegraphics[width=0.45\textwidth]{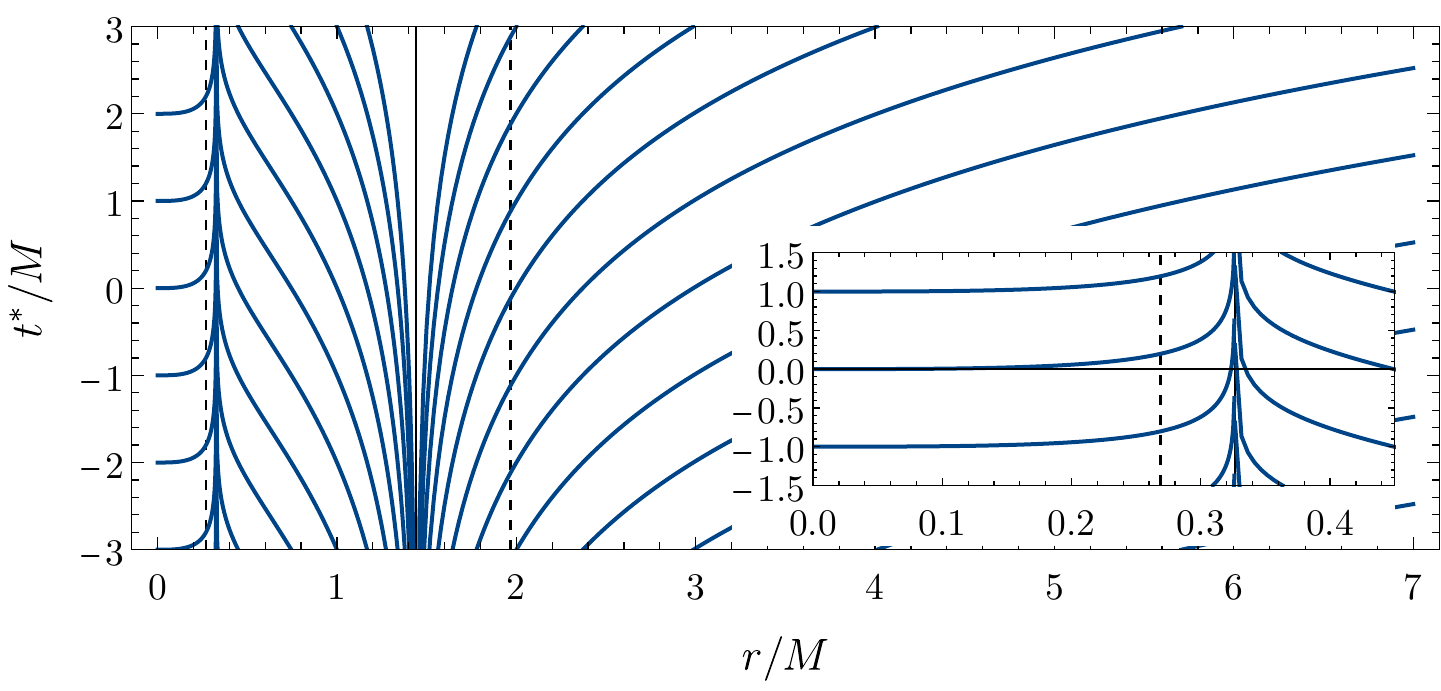}}
    \\
    \subfloat[$\ell = 0.6M$.]{\includegraphics[width=0.45\textwidth]{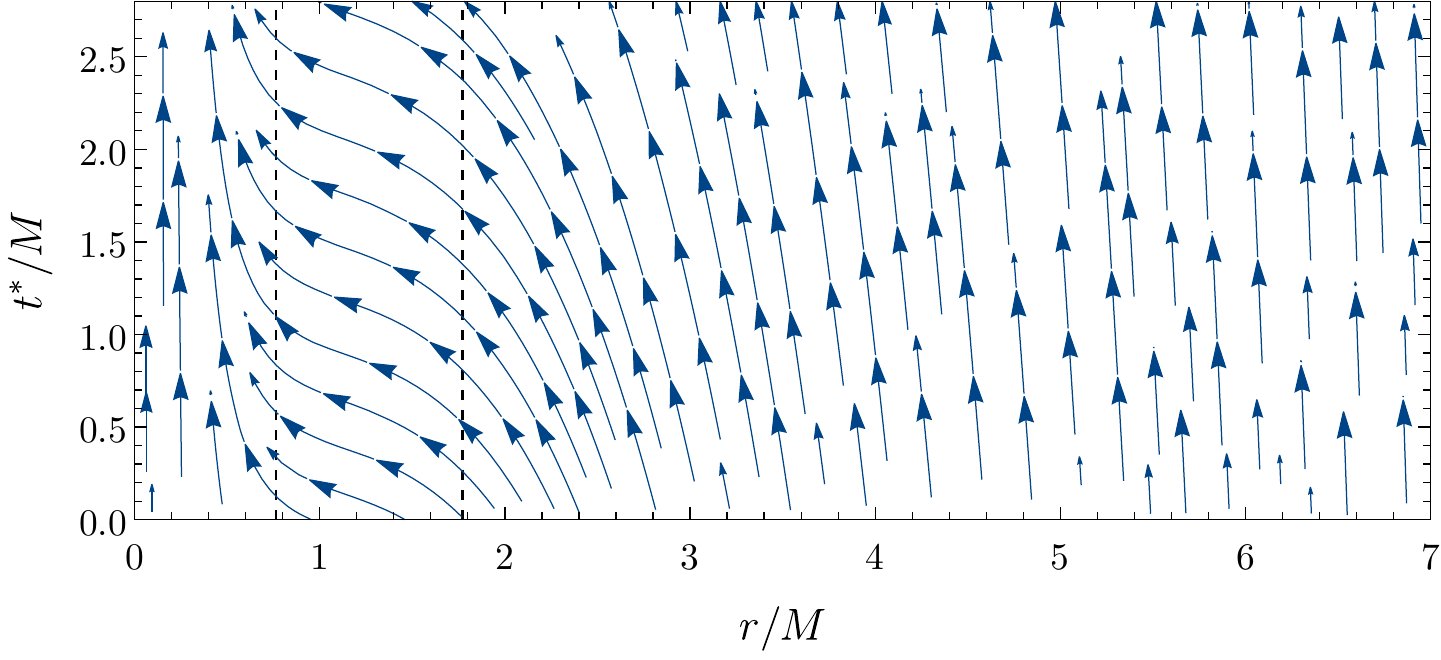}}
    \hfill
    \subfloat[$\ell = 0.6M$.]{\includegraphics[width=0.45\textwidth]{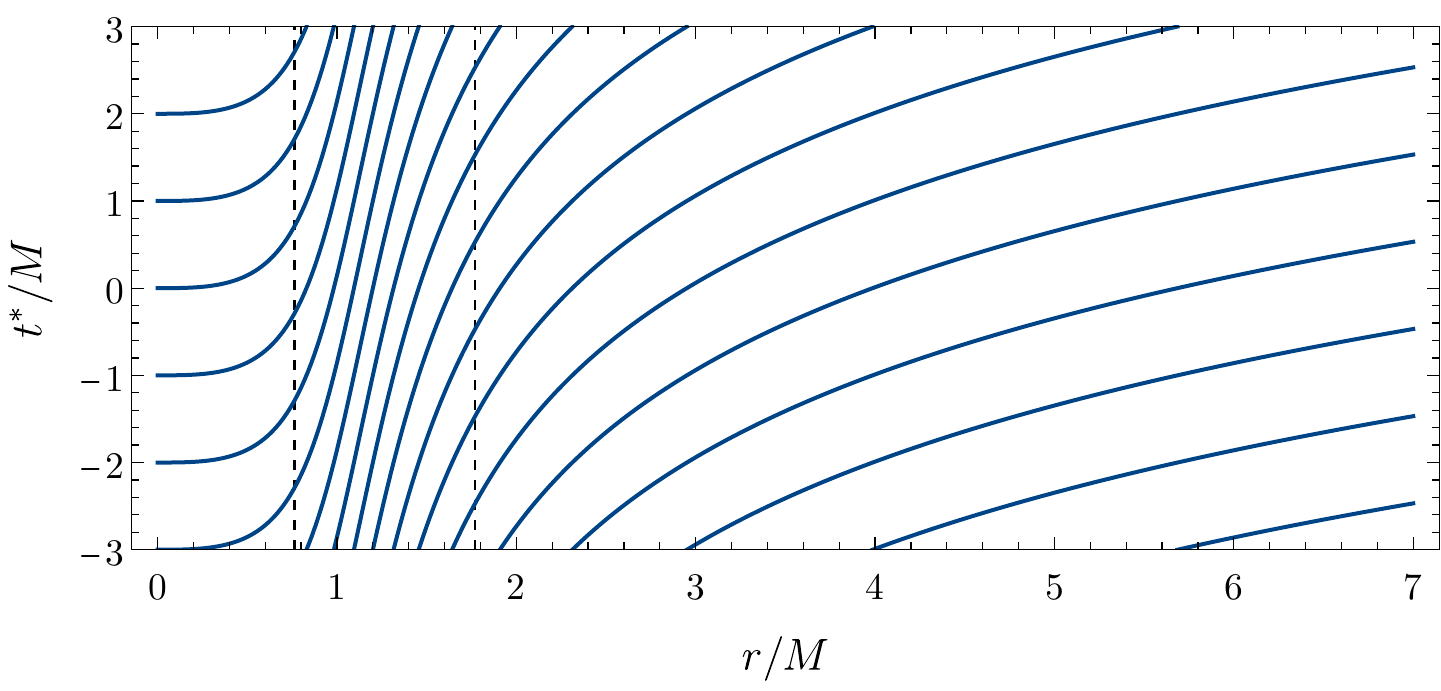}}\\
    \subfloat[$\ell = 1.3M$.]{\includegraphics[width=0.45\textwidth]{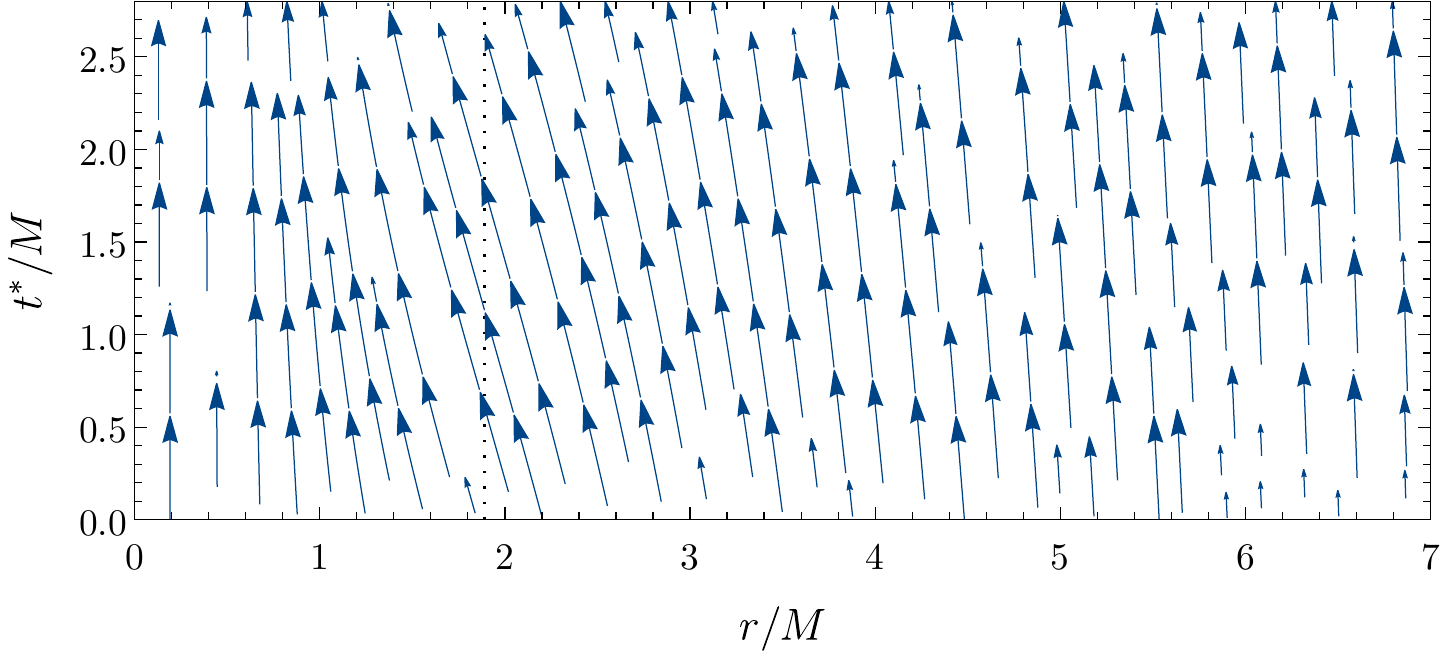}}
     \hfill
    \subfloat[$\ell = 1.3M$.]{\includegraphics[width=0.45\textwidth]{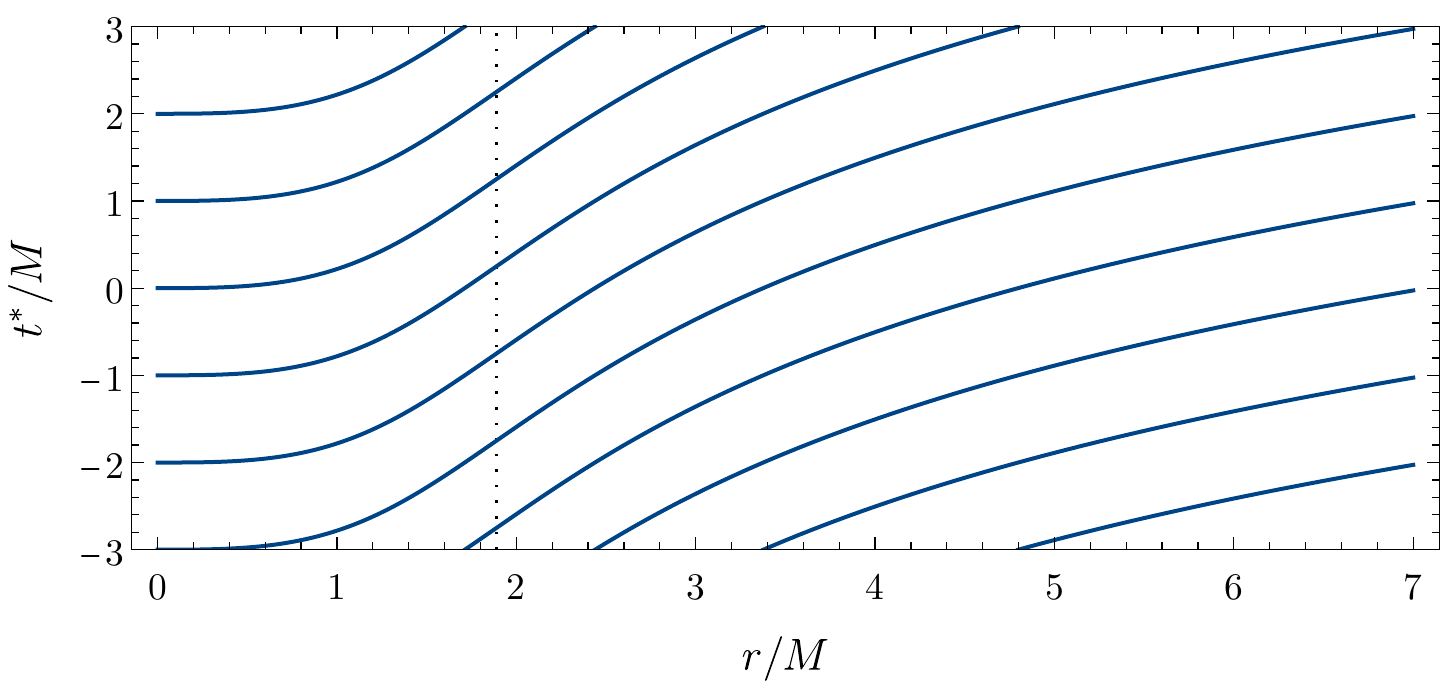}}
    \caption{Hayward non-singular black hole: \aet~flow (left) and constant-khronon surfaces (right). Black solid lines mark universal horizons, dashed lines Killing horizons; the dotted line signals the star's effective radius. The first row depicts the case with outer and inner KHs and UHs, the middle row the case with only two KHs, the bottom row the case of an ultracompact, horizonless, object. The insets are a zoom-in to the small-$r$ region.}
    \label{fig:aekHayward}
\end{figure}

The causal structure corresponding to the Hayward-like regularisation is summarised in \cref{fig:aekHayward}. Each row corresponds to a different value of $\ell$ and therefore to a different regime: the top row represents a non-singular black hole, the second row the intermediate regime and the third row the star-like regime.

The left-hand panel displays the flow of $u_a$, written in the $(t_*, r)$ coordinates: the horizontal component of the arrows is $u_r=y$ while the vertical one is $u_{t_*} = u_a \ \chi^a$. 
The right-hand panel, instead, presents constant-khronon lines.

At large $r$, the \aet~is almost vertical, i.e.~aligned with the Killing vector, and constant-khronon lines are also lines of constant $t_*$. 
As one approaches to smaller $r$, however, the \aet~tilts inwards. 
Nothing remarkable happens at the outer KH, whose location is depicted for reference only. 
At the outer UH, instead, the \aet~is horizontal while the constant-khronon lines exponentially recede away from the UH --- i.e.~they peal off, in an amount that is controlled by the surface gravity, exactly as null rays would do at a trapping horizon. 
Note that behind the outer UH the \aet~points downwards, meaning that it flows in the opposite direction with respect to the Killing vector.

Moving to yet smaller $r$, the \aet~becomes horizontal again at the inner UH. 
The constant-khronon lines instead pile up exponentially at the UH --- as null rays would do at an inner trapping horizon. 
Inside the inner UH the \aet~points again in the same direction as the Killing vector. 
Note that nothing remarkable takes place at the inner KH. 
Close to $r=0$, the flow becomes identical to that of large $r$, i.e.~to that of flat space.

\subsection{Non-connected regularisation}

\begin{figure}[htb!]
    \centering
    \subfloat[$\ell = 0.25M$.]{\includegraphics[width=0.45\textwidth]{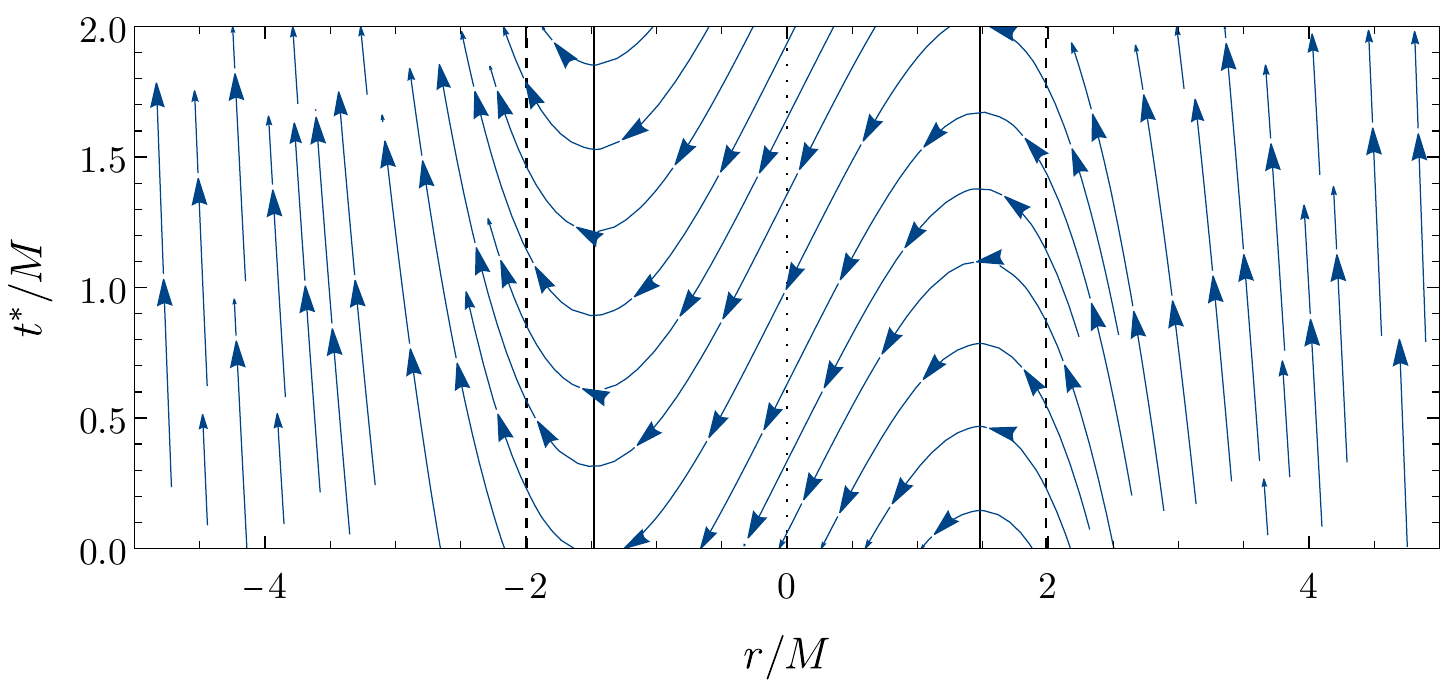}}
    \hfill
    \subfloat[$\ell = 0.25M$.]{\includegraphics[width=0.45\textwidth]{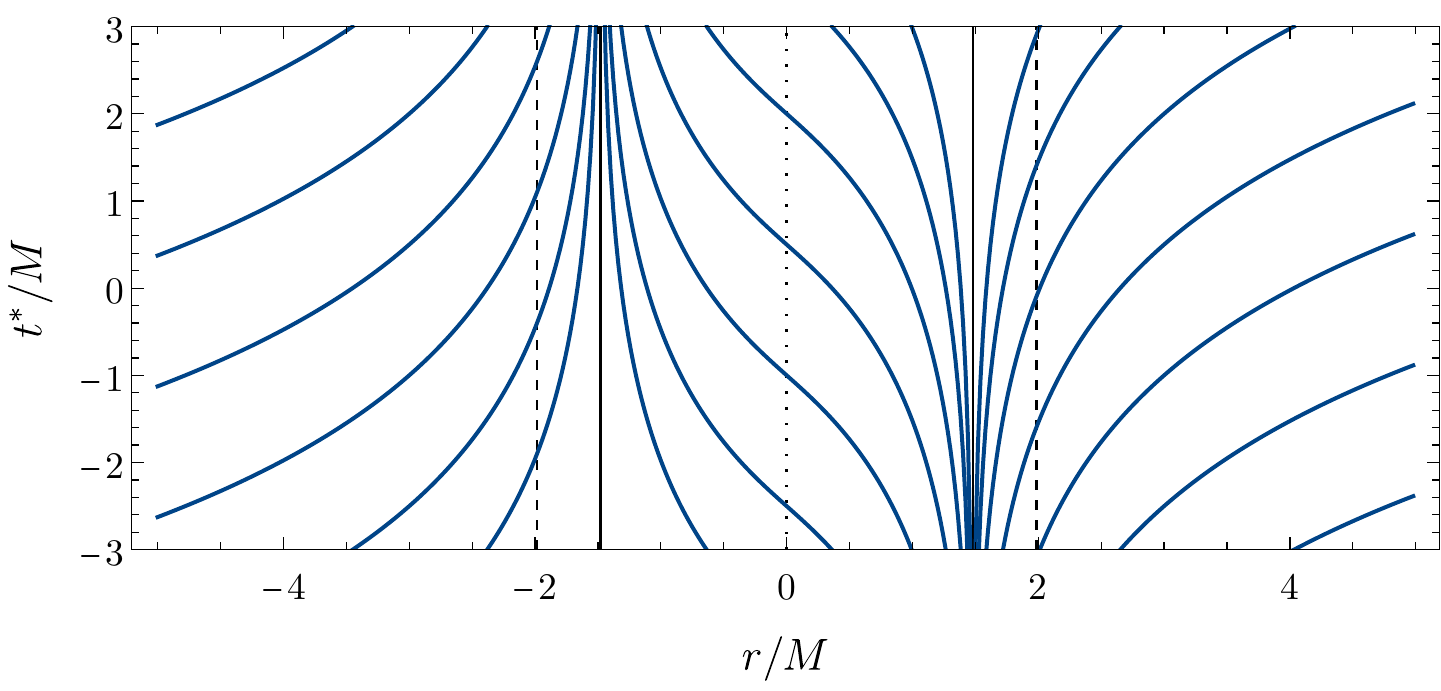}}
    \\
    \subfloat[$\ell = 1.8M$.]{\includegraphics[width=0.45\textwidth]{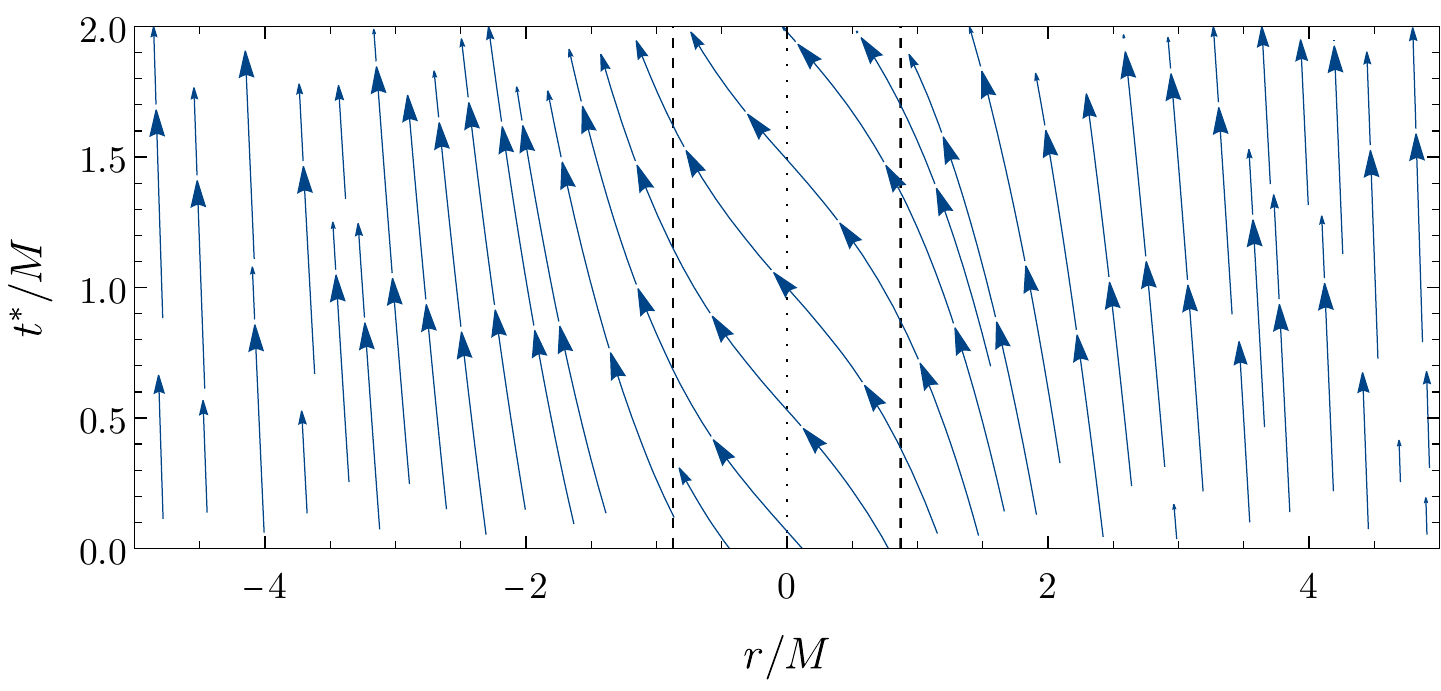}}
     \hfill
    \subfloat[$\ell = 1.8M$.]{\includegraphics[width=0.45\textwidth]{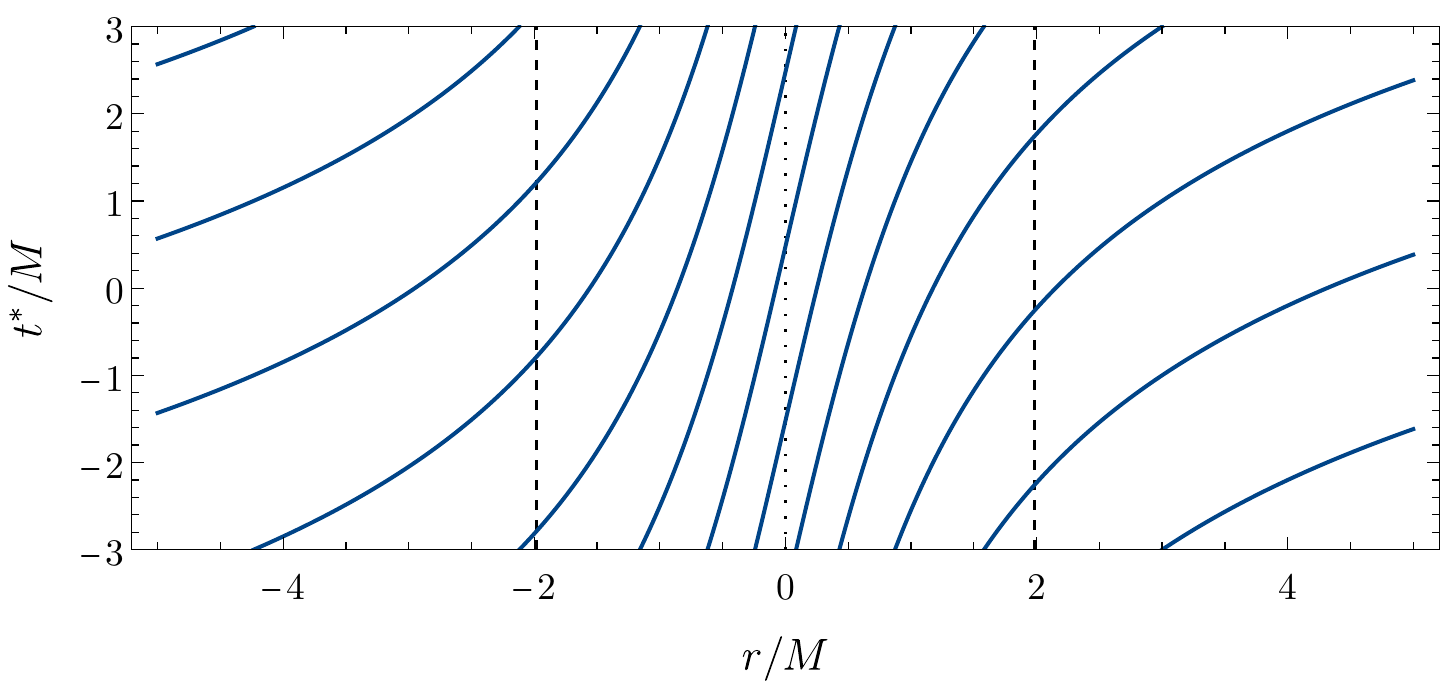}}\\
    \subfloat[$\ell = 2.5M$.]{\includegraphics[width=0.45\textwidth]{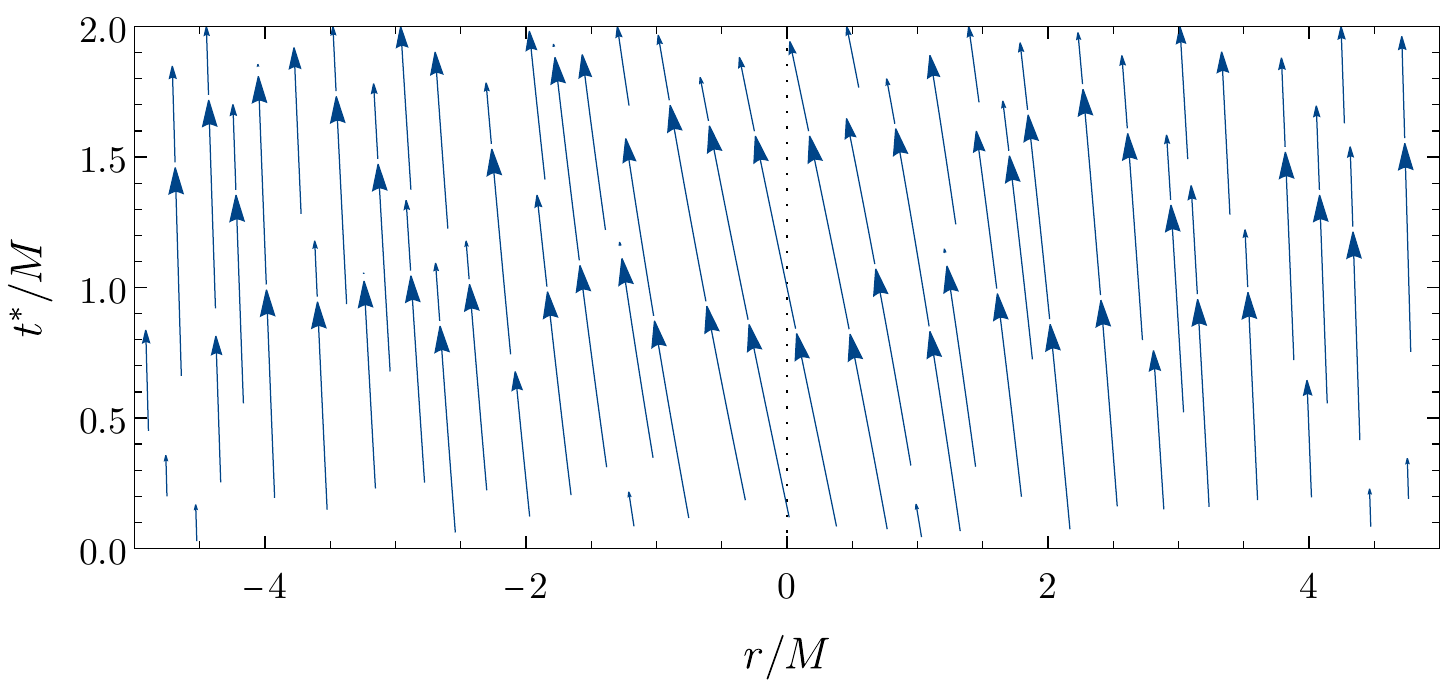}}
     \hfill
    \subfloat[$\ell = 2.5M$.]{\includegraphics[width=0.45\textwidth]{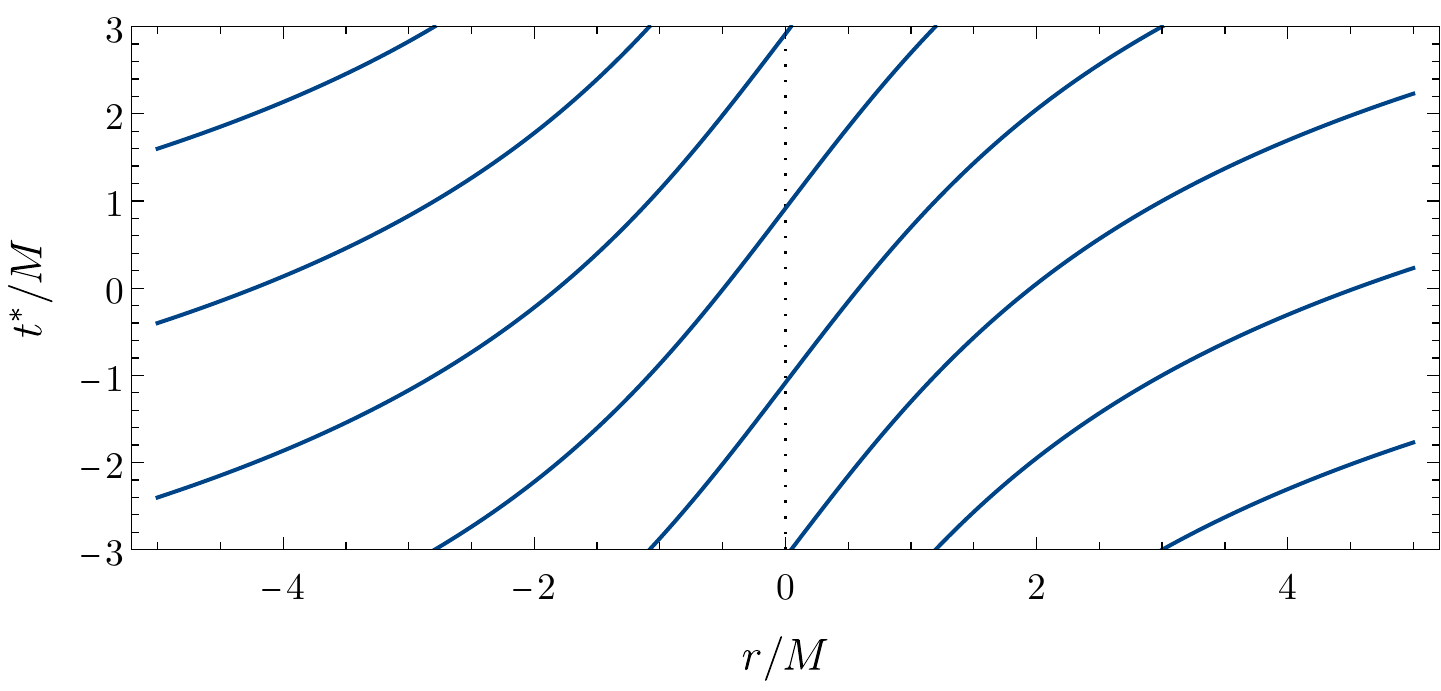}}
    \caption{Black bounce: \aet~flow (left) and constant-khronon surfaces (right). Black solid lines mark universal horizons, dashed lines Killing horizons; the dotted line signals the mouth. the first row depicts a black bounce with a UH and a KH per side, the second row represents a configuration with just one with KHs per side but no UHs, the third row portraits a naked (without horizons) traversable wormhole}
    \label{fig:aekSV}
\end{figure}

The \aet~flow and constant-khronon lines for the black-bounce-like regularisation are displayed in \cref{fig:aekSV}. 
As before, each row corresponds to a different regime: the first to a black bounce with UHs, the second to one with KHs but no UHs and the third to one without horizons.

The \aet, which is aligned with the Killing vector at infinity, tilts inwards as one moves closer to $r=0$. 
At the UH it becomes horizontal, it flows in the opposite direction until it becomes horizontal again at the UH in the other universe and then returns aligned with the Killing vector at $r=-\infty$. 
The constant-khronon curves pile off the UH in our universe and pile up at the UH in the other universe. 
Note that the KHs and the throat look like any other location to the \aet.

\section{Deviations from vacuum, or The effective SET}\label{sec:matter}

As already mentioned in the Introduction, \cref{sec:Introduction}, the non-singular configurations we are describing are supposed to be solutions of the full Ho\v{r}ava action, and in this sense cannot be expected to be vacuum solutions of its low-energy version --- khronometric theory. 

Nevertheless, it is still instructive to study the extent to which our configurations fail to be (vacuum) solutions.
Our strategy is inspired by the analysis of non-vacuum solutions, since any metric and \aet~flow can be seen as solutions of the equations of motion with an appropriate matter content (in this case, one that couples to the metric and to the \aet). 
For this reason, we will speak of ``effective sources'' and use terms such as ``energy density'' and ``pressures''. 
It should be clear, however, that this is merely a choice of terminology and we are not positing the existence of any physical form of matter.  
Indeed, such density and pressures can be seen as components of an effective SET induced by the higher-order terms of the Ho\v{r}ava action. 
Understanding whether this is actually the case is a crucial but open question, given the daunting task of manipulating such extra terms. 

Thus, in the following we will characterise the form and distribution of such effective sources, in order to better understand the regular geometries we are proposing. 
In particular, we will show that these sources are sizeable only in the proximity of the would-be singularity and decay very rapidly as one moves to larger radii. For all practical purposes, therefore, the spacetime surrounding our regular objects can be considered vacuum and their phenomenology can thus be studied using standard tools of general relativity.

Let us start by noticing that the equation of motion for the khronon \cref{eq:eomAE} can be written as
\be
\frac{1}{N} J = 0 \qq{with}  J = \left[ \nabla_a \mathcal{A}^a - a_a \mathcal{A}^a \right]\, .
\ee
The lapse $N$ can be chosen (almost) arbitrarily, since it depends on how the khronon is parametrised; the scalar quantity $J$ instead only depends on the \aet~and can be computed unequivocally. Thus, $J$ will be the khronon's effective source.

Similar considerations hold for the Einstein's equations \cref{eq:eomg}. Since the components of $\mathcal{G}_{ab}$ clearly depend on the choice of coordinates, one first needs to find a coordinate-independent way of characterising the source.
One way to achieve this goal would be to compute its eigenvalues, which are scalars under general coordinate transformations. When the eigenvalues are real, they can be interpreted as energy density and principal pressures of some non-perfect fluid.
Unfortunately, while this characterisation works for the Einstein's tensor, it fails for the SET of the \aet, since there are regions in the spacetime where the eigenvalues are complex (i.e.~the \aet's SET is of Type IV in the Hawking--Ellis classification~\cite{hawking_large_1973}). 

However, since in the framework of Ho\v{r}ava gravity there exists a preferred foliation and therefore a preferred observer, it makes sense to characterise the deviations from vacuum as measured by such observer. Hence, we compute the projection
\be
\rho^{(u)} = \mathcal{G}_{ab} u^a u^b\, ,
\ee
which we could interpret as the energy density measured by an observer that is comoving with the \aet.
We then pick another vector $s_a$ that is spacelike, outward-pointing, of unit norm and orthogonal to $u_a$ and use it to define a radial pressure as 
\be
p^{(s)} = \mathcal{G}_{ab}s^as^b\, .
\ee
We use
\be\label{eq:s}
s^a \partial_a = A \partial_v + w \partial_r \qq{with} w = +\frac{1+A^2F}{2A}\, .
\ee
Finally, we could define a tangential pressure in an analogous way, or simply as
\be
p^\perp = -\mathcal{G}^\theta_{ \ \theta} = -\mathcal{G}^\phi_{ \ \phi}\, .
\ee

Since the parameters $\alpha,\ \beta,\ \lambda$ enter the action as coupling constants, all the scalars that we have just introduced share the same simple structure. Consider $\rho^{(u)}$ as an example: it can be written as
\be
\rho^{(u)} = \rho^{(u)}_G + \alpha \rho^{(u)}_\alpha + \beta \rho^{(u)}_\beta + \lambda \rho^{(u)}_\lambda\, .
\ee
Here, $\rho^{(u)}_G$ derives from the Einstein's tensor while each of $\rho^{(u)}_\alpha,\ \rho^{(u)}_\beta,\ \rho^{(u)}_\lambda$ derives from the operators that appear in the action multiplied respectively by $\alpha,\ \beta$ and $\lambda$. Clearly, each of them still depends on $\beta$ (and on $\ell$), since the explicit form of the metric and of the \aet~does; but not on $\alpha$ nor $\lambda$. 
We will use analogous notations for the decompositions of $p^{(s)},\ p^\perp$ and $J$, with the only difference that $J$ has no ``$J_G$'' part. 
One can check that $p^{(s)}_\alpha = -p^\perp_\alpha$ in all the cases that we consider.

A remark is in order, at this point. 
The singular geometry of \cref{eq:exact_met,eq:exact_A} is a solution of the equations of motion only for $\alpha=0$.
Indeed, as will be made explicit in the next two subsections, the effective sources proportional to $\alpha$ generically do not vanish --- not even in the limit $\ell \to 0$, in which the singular geometry is retrieved.
Hence, one might worry that allowing $\alpha \neq 0$ in the analysis of the non-singular configurations is not consistent.

Here, however, we choose to keep $\alpha \neq 0$. 
The reason is that, in the absence of some custodial symmetry that protects it against running, the higher-order operators in the action of Ho\v{r}ava gravity will generically affect the value of $\alpha$ (as well as that of $\beta$ and $\lambda$) at the level of the effective field theory.
Hence, we cannot presume it to be zero at this stage.

Still, at low energies, the effect of higher-order operators is negligible and the value of $\alpha$ is the one set by (low-energy) observations: $\alpha \lesssim \order{10^{-4}}$ --- cf.~\cref{sec:theo}. 
One should bear in mind, therefore, that at large distances the effective sources proportional to $\alpha$ are \emph{highly} suppressed.

\subsection{Hayward's effective sources}

Here, we remain agnostic on the specific choice of $r_0(r)$ for as long as possible; however, the explicit results are often cumbersome and not particularly enlightening.
For this reason, we specialise to Hayward's choice and discuss, in particular, the asymptotic behaviour of the effective sources.

\paragraph{Khronon's equation} One finds that $J_\beta = J_\lambda$; clearly, these are zero when $\ell = 0$. 
$J_\alpha$ instead is non-zero even in the limit in which the regularisation parameter vanishes, since the singular solution we started with is an exact (vacuum) solution only for $\alpha =0$.
The explicit expressions are not particularly enlightening and we hence omit them here. We can however get useful insights by looking at their asymptotic behaviour.

At infinity, we find
\be
J &=\alpha J_\alpha+(\beta+\lambda) J_{\beta,\lambda} \nonumber \\ 
&=\alpha \left[ -6\sqrt{3}\sqrt{\frac{1}{1-\beta}} \frac{M^4}{r^5} + \order{r^{-6}} \right] + (\beta + \lambda) \left[ 540\sqrt{3} \sqrt{\frac{1}{1-\beta}} \frac{M^3 \ell^2}{r^6} + \order{r^{-7}} \right] \, .
\ee
The different scaling between $J_\alpha$ and $J_{\beta,\lambda}$ is not surprising, since the former does not vanish in the $\ell \to 0$ limit --- as previously argued. 
In any case, it is easy to see from the above expression that the effective source vanishes rapidly as one moves away from the object. 

The sources may be large at intermediate radii, but become very small in the opposite limit, small $r$, and vanish exactly at $r=0$.
Indeed, expanding around this point, we find
\be
J = \alpha \left[ \frac{27 \sqrt{3}}{4} \sqrt{\frac{1}{1-\beta}} \frac{r^5}{\ell^6} + \order{r^6} \right] + (\beta + \lambda) \left[27 \sqrt{3} \sqrt{\frac{1}{1-\beta}} \frac{r^3}{\ell^4} + \order{r^4} \right]\, .
\ee

Hence, the connected non-singular configuration is almost a solution of khronometric theory at very large and very small distances from the centre.

\paragraph{Einstein's equations} Plugging in the Ans\"{a}tze \cref{eq:regConm,eq:regConae}, we find a series of additional identities:
\be
\rho ^{(u)}_G = - p^{(s)}_G \, , & \quad p^\perp_\lambda = p^{(s)}_\lambda \\
\rho^{(u)}_G + \beta \rho^{(u)}_\beta = \frac{r_0'}{r^2} + \beta \rho^{(u)}_\lambda\, ,  \quad p^{(s)}_G + \beta p^{(s)}_\beta &= -\frac{r_0'}{r^2} + \beta p^{(s)}_\lambda\, , \quad p^\perp_G + \beta p^\perp_\beta = -\frac{r_0''}{2r} + \beta p^\perp_\lambda
\ee
Hence, we can write
\be
\rho^{(u)} &= \frac{r_0'}{r^2} + (\beta+\lambda) \frac{27  r_0^2}{128(1-\beta)}\left[ \frac{r_0'}{r^2} \right]^2 + \alpha \rho^{(u)}_\alpha  \, ,\\
p^{(s)} &= -\frac{r_0'}{r^2} - (\beta+\lambda) \frac{27 r_0^2}{128(1-\beta) r^5}\left[ 2r (r_0')^2 + r_0(r r_0'' - 2r_0')\right] + \alpha p^{(s)}_\alpha\, , \\
p^ \perp &= -\frac{r_0''}{2r} - (\beta+\lambda) \frac{27 r_0^2}{128(1-\beta) r^5}\left[ 2r (r_0')^2 + r_0(r r_0'' - 2r_0')\right] - \alpha p^{(s)}_\alpha
\ee
The expressions of $\rho^{(u)}_\alpha$ and $p^{(s)}_\alpha$ are slightly more involved and we omit them here.

Focusing on Hayward's choice, we can read off the asymptotic behaviours.
At large $r$ we have
\be
\rho^{(u)} &= \left[ \frac{12M^2\ell^2}{r^6} + \order{r^{-9}} \right] + (\beta +\lambda) \left[ \frac{243M^6 \ell^4}{2(1-\beta) r^{12}} + \order{r^{-15}} \right] + \alpha \left[ \frac{M^2}{2r^4} + \order{r^{-5}} \right] \, , \\
p^{(s)} &= - \left[ \frac{12M^2\ell^2}{r^6} + \order{r^{-9}} \right]  + (\beta + \lambda) \left[ \frac{243M^5\ell^2}{2(1-\beta) r^9} + \order{r^{-12}}\right] + \alpha \left[ \frac{M^2}{2r^4} + \order{r^{-5}} \right] \, ,\\
p^\perp &= \left[ \frac{24 M^2 \ell^2}{r^6} + \order{r^{-9}}\right]  + (\beta+\lambda) \left[ \frac{243M^5 \ell^2}{2(1-\beta) r^9} + \order{r^{-12}}\right] - \alpha \left[ \frac{M^2}{2 r^4} + \order{r^{-5}} \right] \, .
\ee
Clearly, these effective sources display ``tails'' that extend, in principle, up to infinity; the tails however decrease very rapidly, hence for all practical purposes the spacetime surrounding the object can be considered empty.
Further note that, as anticipated, these sources do not vanish in the limit $\ell \to 0$ because of the terms $\propto \alpha$. 
This is simply due to the fact that the singular configuration is a solution only for $\alpha =0$.

At $r=0$, the following expansions hold:
\be
\rho^{(u)} &= \left[ \frac{3}{\ell^2} + \order{r^3} \right] + (\beta + \lambda) \left[\frac{243 r^6}{128(1-\beta) \ell^8}  + \order{r^7} \right]  + \alpha \left[ \frac{3}{\ell^2} + \order{r} \right] \, , \\
p^{(s)} &= - \left[ \frac{3}{\ell^2} + \order{r^3} \right] - (\beta +\lambda) \left[\frac{243 r^6}{64(1-\beta) \ell^8} + \order{r^7}\right]  + \alpha \left[ \frac{r^2}{2\ell^4} + \order{r^3} \right] \, , \\
p^\perp &= - \left[ \frac{3}{\ell^2} + \order{r^3}\right]  - (\beta + \lambda) \left[ \frac{243 r^6}{64(1-\beta) \ell^8}  + \order{r^7} \right] - \alpha \left[ \frac{r^2}{2\ell^4} + \order{r^3} \right] \, .
\ee
Note that the Einstein's tensor presents a de Sitter form.

\begin{figure}[t]
    \centering
    \subfloat[\label{fig:matterH_G} Energy density and principal pressures deriving from the Einstein's tensor.]{\includegraphics[width=0.48\textwidth]{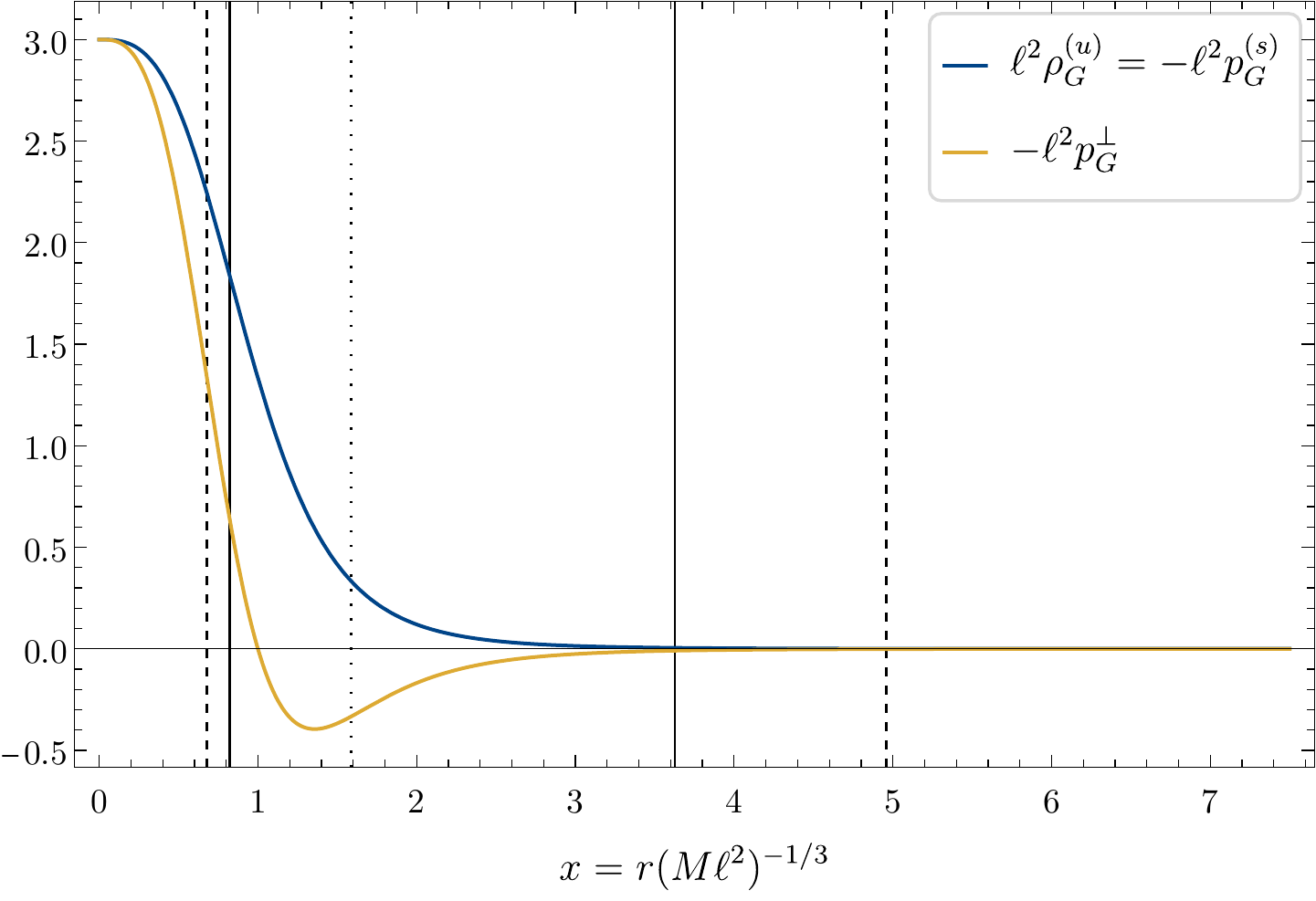}}
    \hfill
    \subfloat[\label{fig:matterH_l} Energy density and principal pressures derived from the \aet's SEMT. These curves should be multiplied by $\lambda$.]{\includegraphics[width=0.48\textwidth]{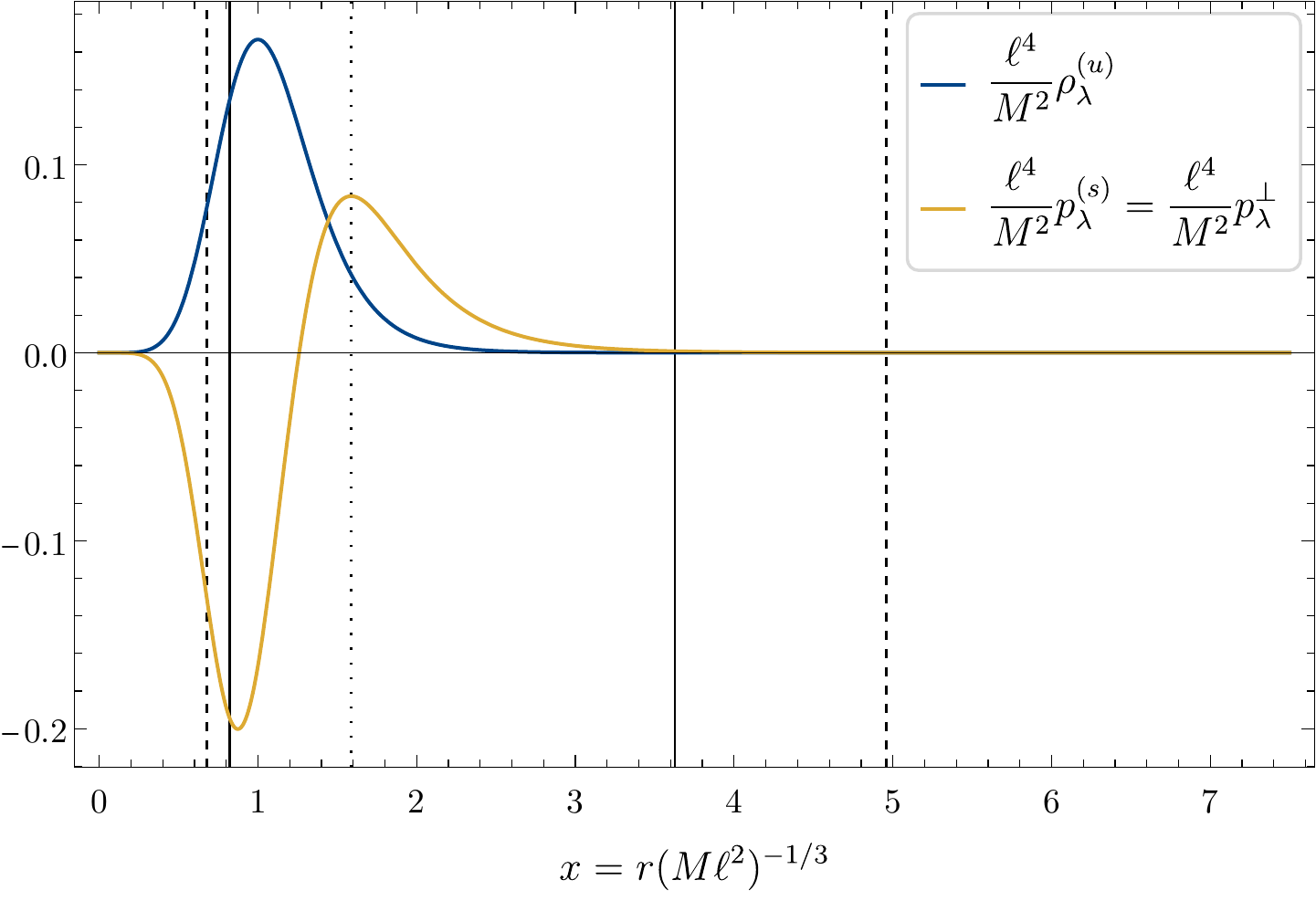}}
    \caption{Components of the energy density and the principal pressures, as measured by an observer comoving with the \aet, for the Hayward non-singular configuration in the minimal theory $\alpha = \beta = 0$. The position of the horizons in this coordinate strongly depends on $\ell$: the vertical black lines mark UH (solid) and KH (dashed) for $\ell = 0.25 M$; the dotted line corresponds to the star's effective radius (which coincides with the degenerate KH in the extremal case).}
    \label{fig:matterH}
\end{figure}

Focusing on the minimal theory ($\alpha=\beta=0$), the analytic expressions become more tractable. We report them for completeness:
\be
\rho^{(u)}_G = -p^{(s)}_G = \frac{12M^2\ell^2}{ \left( r^3 + 12M \ell^2 \right)^2}\, , & \quad p^\perp_G  = -\frac{24M^2\ell^2 (M \ell^2-r^3)}{(r^3 + 2M \ell^2)^3} \, ,\\
\rho^{(u)}_\lambda = \frac{243 M^6 \ell^4 r^6}{2(r^3 + 2M\ell^2)^6} \, ,& \quad p^{(s)}_\lambda = p^\perp_\lambda =  \frac{243 M^5 \ell^2 r^6 (r^3 - 2M\ell^2)}{2(r^3 + 2M\ell^2)^6}\, ;
\ee
and provide their plots in \cref{fig:matterH}. 

In order to produce the figures, we exploit a self-similarity property that these functions enjoy: when written in terms of the variable $x = r \left( M \ell^2 \right)^{-1/3}$, they only depend on $\ell$ through a multiplicative factor, which we can remove.
Thus, the curves in \cref{fig:matterH} are $\ell$-independent: the $\ell$-dependence can be reinstated by simply rescaling the axes appropriately.

The location of the horizons in the $x$ coordinate depends markedly on $\ell$: for reference, \cref{fig:matterH} reports an illustrative example.
It also reports the location of the star's effective radius, which is defined only for $\ell > 4M /(3\sqrt{3})$ but has an otherwise $\ell$-independent $x$-coordinate: 
\be
r_\star =  \left( 4M \ell^2 \right)^{1/3} \mapsto x_\star = 4^{1/3} \simeq 1.59\, .
\ee
The fact that $x_\star$ does not depend on the regularisation parameter might sound suspicious, as it seems to suggest that the value of the effective sources at the star's radius is always the same, irrespective of the value of $\ell$.
Physical intuition would suggest the opposite: the sources corresponding to large stars should be ``dilute'' with respect to those of more compact stars.
Physical intuition does indeed paint the correct picture: the value of each of the effective sources at the star's radius is given by a constant (of order one in appropriate units) \emph{divided by $\ell^2$}. 
So increasing $\ell$ does suppress the deviations away from vacuum. 

Inspecting the figure, we realise that the effective sources are typically negligible at the scale of the outer horizon. 
They are still small, though less so, at the scale of the star's radius, and become almost zero very rapidly as one moves outwards.
We deduce that most of the phenomenologically relevant phenomena involving these non-singular objects take place, for all practical purposes, in vacuum.

\subsection{Black bounce's effective sources}

In the non-connected case, we are forced to analyse the specific example provided by the black bounce spacetime. 
The analytic expressions are often reasonably compact: when this is the case, we report them in full. 
However, as in the previous section, we will put the emphasis on the more informative asymptotic behaviours of the effective sources. 

\paragraph{Khronon's equation} We find that $J_\lambda = 0$ identically. This means that, remarkably, the khronon's equation of motion is satisfied in the minimal theory $\alpha=\beta=0$. For the more general cases,  $J_\alpha$ and $J_\beta$ can be written as
\be\label{eq:JSV}
J_\alpha = r \left[M \varrho^2 P_5(\varrho) + \ell^2 P_6(\varrho) \right] j_\alpha(\varrho) \qq{and} J_\beta = r \ell^2  j_\beta(\varrho)\, ,
\ee
where $j_\alpha$ and $j_\beta$ do not depend on $\ell$ while $P_5 (\varrho) $ and $P_6(\varrho)$ are polynomials of degree five and six, respectively, in $\varrho$.

At infinity, one finds the following expansions:
\be
J = \alpha \left[ -6\sqrt{3}\sqrt{\frac{\beta}{1-\beta}} \frac{M^4}{\varrho^5} + \order{\varrho^{-6}} \right] + \beta \left[12 \sqrt{3} \sqrt{\frac{\beta}{1-\beta}} \frac{M^2 \ell^2}{\varrho^5} + \order{\varrho^{-6}} \right]\, ,
\ee
so even in this case the effective sources go to zero very rapidly.

As the expressions in \cref{eq:JSV} make explicit, these functions are $\order{r}$ close to $r=0$, for all nonzero values of $\ell$. Note that the limit $\ell \to 0$ is, as expected, singular.

\paragraph{Einstein's equations} We find that the term proportional to $\lambda$ in $\mathcal{G}_{ab}$ vanishes identically: this means that, in the minimal theory $\alpha = \beta =0$, the only deviations from vacuum come from the Einstein tensor. 
In other words
\be
\rho^{(u)}_\lambda = p^{(s)}_\lambda = p^\perp_\lambda = 0\, .
\ee

The analytic expression of the effective energy density is
\be
\rho^{(u)} = - \frac{\ell^2 }{8\varrho^8} \left( 8 \varrho^4 - 32 \varrho^3 M + 27 M^4 \right) + \alpha \left[M  \frac{\varrho^2 M P_4(\varrho) + \ell^2 P_5(\varrho)}{8\varrho^8 \left( 4\varrho^2+4\varrho M+3M^2 \right)} \right]\, ,
\ee
where the $P_n$ are polynomials of degree $n$ in $\varrho$; note in particular that nothing depends on $\beta$ --- $\rho^{(u)}_G$ and $\rho^{(u)}_\beta$ separately do, but the sum $\rho^{(u)}_G+\beta \rho^{(u)}_\beta$ does not. 
For the pressures, we find
\be
p^{(s)} &= -\frac{\ell^2}{8\varrho^8} \left( 8\varrho^4 + 27 M^4 \right) + \alpha \left[ \frac{M^2 r^2 \left(4\varrho^2 + 6\varrho M + 9M^2 \right)^2}{8\varrho^8 \left( 4\varrho^2+4\varrho M+3M^2 \right)}\right]\, ,\\
p^\perp &=  \frac{\ell^2(\varrho-M)}{\varrho^5} - \alpha \left[ \frac{M^2 r^2 \left(4\varrho^2 + 6\varrho M + 9M^2 \right)^2}{8\varrho^8 \left( 4\varrho^2+4\varrho M+3M^2 \right)}\right]\, ;
\ee
as before, the $\beta$-dependence cancels out.

Once again, we focus on the asymptotic behaviour. At infinity 
\be
\rho^{(u)} &= \left[ - \frac{\ell^2}{\varrho^4} + \order{\varrho^{-5}}\right] + \alpha \left[ -\frac{M^2}{2\varrho^4} + \order{\varrho^{-5}} \right]\, , \\
p^{(s)} &= -p^\perp + \order{\varrho^{-5}}  = \rho^{(u)} + \order{\varrho^{-5}}  \, . 
\ee

\noindent Note that the fall-off rate of the tails is still rather fast. 
Again, it is easy to see that even for $\ell\to 0$ one does not recover vacuum if $\alpha\neq 0$. As in the previous case this is simply due to the fact that only for $\alpha=0$ the considered singular BH spacetime is an exact solution of the field equations in vacuum.

At $r=0$, instead, the values of the sources are nonzero and controlled by the regularisation parameter $\ell$:
\be
\rho^{(u)} &= - \frac{8 \ell^4 +27M^4 - 32\ell^3 M }{8\ell^6} + \alpha \left[ \frac{27M^4 - 8M \ell^3}{8\ell^6} \right] \, , \\
p^{(s)} &= - \frac{8\ell^4+27M^4}{\ell^6} \, ,\\
p^\perp &= \frac{\ell - M}{\ell^3} \, .
\ee
For symmetry reasons, the throat is an extremal point (either a local minimum or maximum) for these functions.

\begin{figure}
    \centering
    \includegraphics[width=.7\textwidth]{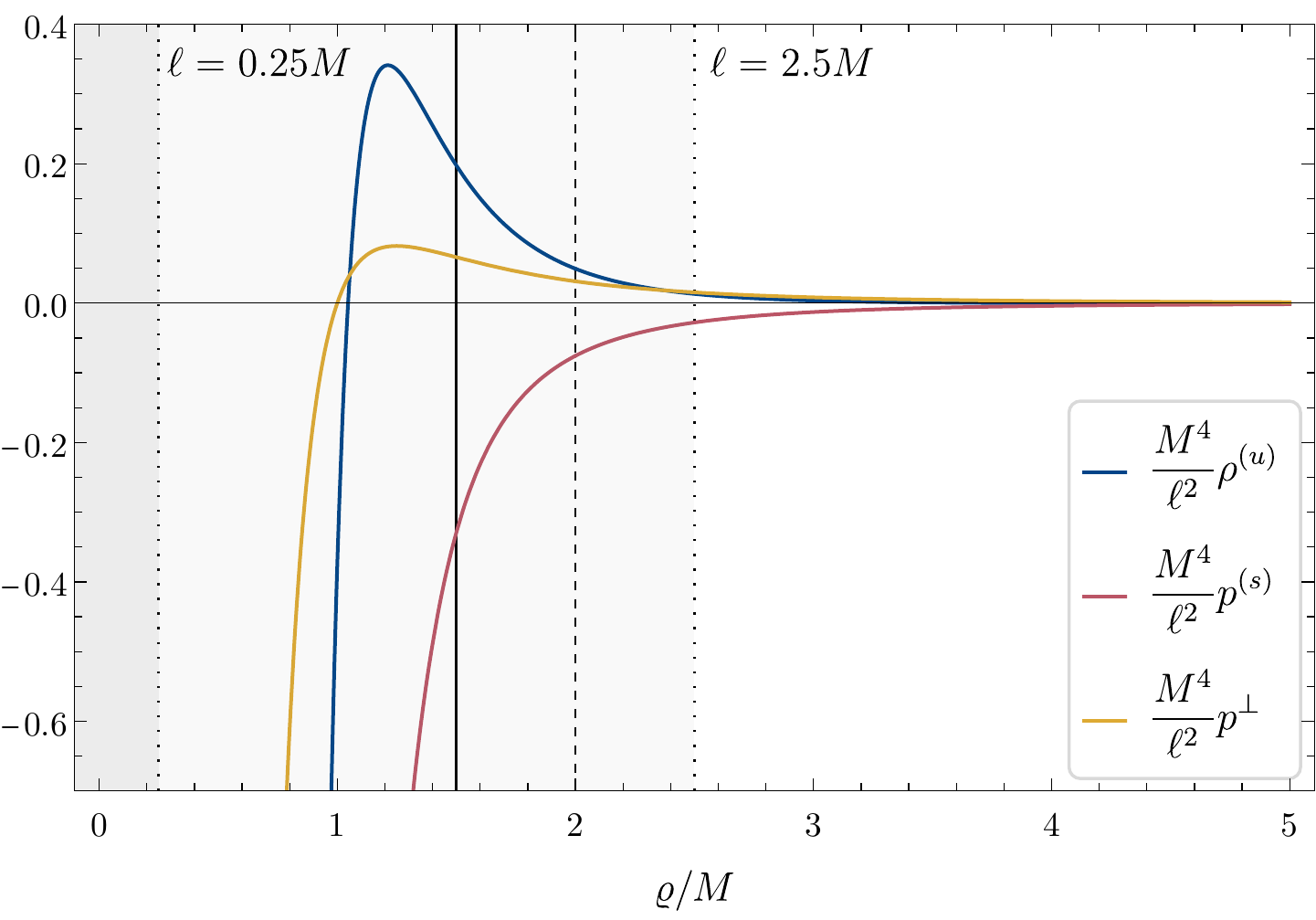}
    \caption{Energy density and principal pressures, as measured by an observer comoving with the \aet, for the black-bounce non-singular metric in the minimal theory $\alpha = \beta = 0$. The black vertical lines mark the UH (solid) and KH (dashed), which may be present or not depending on the value of $\ell$. Dotted lines signal the position of the mouth for two choices of $\ell$, corresponding to a hidden ($\ell = 0.25 M$) and a traversable ($\ell=2.5 M$) wormhole respectively. Recall that, since $\varrho = \sqrt{r^2+\ell^2}$, the region $\varrho<\ell$ is unphysical and should be removed; for this reason it is shaded.
    }
    \label{fig:matterSV}
\end{figure}

Finally, we again focus on the minimal theory and provide plots of the nonzero sources. 
As the analytic expressions make clear, once the coordinate $\varrho$ is employed the dependence on $\ell$ is trivial, since this parameter only enters as a multiplicative factor. 
For this reason, we decide to plot, in \cref{fig:matterSV}, the $\ell$-independent part only, as a function of $\varrho$. 
For reference, dotted lines mark the location of the wormhole mouth for two specific choices of $\ell$, corresponding to a hidden and a traversable wormhole respectively. 

We remind the reader that, although the plot extends to $\varrho=0$, $\min(\varrho)=\ell$.
Hence, for any given choice $\ell$, the region $\varrho<\ell$ does not belong to the spacetime and should therefore be removed: in \cref{fig:matterSV} this is rendered by shading.
The curves should thus be cut off at $\varrho=\ell$ and joined smoothly with a mirror copy of themselves; moreover, they should be multiplied by $\ell^2$. 

The upshot of this analysis is that the deviations away from vacuum are sizeable only in a region close to the mouth, but decay very fast as one moves away from the object. 
Therefore, the physics in the surrounding of the black bounce is well described by the equations of vacuum khronometric theory.

\section{Conclusions} \label{sec:conc}

In this paper, we have presented explicit examples of simply connected and non-connected regularisations of a static and spherically symmetric black-hole solution in low-energy Ho\v{r}ava gravity.

In the simplest settings, these regularisations are minimal deformations of the singular solution that only depend on one parameter, the value of which determines whether universal and/or Killing horizons are present. Thus, each of the metrics and \aet~flows that belong to the connected class describes a compact object that is either a regular black hole or an horizonless ``star'' with (anti-)de Sitter core --- depending on the value of said parameter. Similarly, each of the metrics and \aet~flows in the non-connected class corresponds to a wormhole whose mouth may or may not be hidden behind horizons. To our knowledge, our examples constitute the first instances of objects of these kinds in the context of khronometric theory.

Our proposals are summarised in \cref{eq:regConm,eq:regConae} (connected) and \cref{eq:SVmet,eq:SVaet} (non-connected).
They are \emph{not} solutions of khronometric theory; however, they instantiate all of the few physically viable classes of non-singular end states of gravitational collapse: connected regular black holes and horizonless objects, non-connected hidden and traversable wormholes.
Therefore, if non-projectable Ho\v{r}ava gravity is indeed a UV completion of khronometric theory, gravitational collapse in the full theory should produce solutions that are qualitatively akin to the configurations that we have hereby described.

In this frame of mind, the deviations of our configurations from the vacuum of khronometric theory should be interpreted as arising from the contributions of higher-order operators.
Checking directly whether this is the case is probably close to impossible.
It seems important, therefore, to seek indirect ways of validating this conjecture.

Even within the low-energy theory, however, our proposals present several intriguing features.
For instance, when a UH is present both the connected and non-connected configurations represent one-parameter generalisations of the exact solution.
Notably, this additional parameter can be used to trim the surface gravity of the UH, thereby affecting the thermal properties of the black hole.
In particular, it seems possible to construct configurations in which the UH is extremal, i.e.~its surface gravity vanishes.

Moreover, connected non-singular black holes necessarily have multiple UHs. 
Since the peeling properties of the inner UH are highly reminiscent of those of inner KHs in GR, which are believed to be unstable (see e.g.~\cite{carballo-rubio_viability_2018,carballo-rubio_inner_2021,di_filippo_inner_2022} and references therein), it is legitimate to wonder whether inner UHs will be subject to a similar instability. 
Indeed, while the instability we are referring to --- the so called \emph{mass inflation} --- manifests itself as an exponential growth in the amplitude of perturbations close to an inner horizon in relativistic physics, non-relativistic settings like ours usually tame such growth due to the existence of modified dispersion relations which prevent an unbounded accumulation of energy at inner KHs. 
This might not be the case for inner UHs, because of what we said above, and some dynamical behaviour might appear. 
We think this question deserves further investigation in the future.

Finally, we note that, both on theoretical as well as phenomenological grounds, the Lorentz-violating effects in matter should be suppressed by some energy scale higher than the one associated with Lorentz breaking in Ho\v{r}ava gravity (see e.g.~\cite{liberati_scale_2012} or~\cite{liberati_tests_2013} and references therein). 
This implies that for all practical purposes light rays and test particles, typically used for BH phenomenology nowadays, essentially move along geodesics of the metric --- regardless of the specifics of the modified dispersion relation. 

Previous studies have shown that the geodesics of the Hayward (as well as Bardeen, Dymnikova \textit{et similia}) and black bounce spacetimes are parametrically close to those of the Schwarzschild metric. 
In particular, these metrics typically admit an unstable light ring --- whose properties are connected, for instance, to the shape and size of electromagnetic shadows, or to the frequencies of the longest-lived quasinormal modes --- that lies close to the one of Schwarzschild.
(Moreover, in the horizonless regime they can admit another, stable light ring, located inside the unstable light ring or at the wormhole mouth, which might be associated to a non-linear instability \cite{cunha_light_2017}.) 

Therefore, when probed at low energies --- e.g.~employing very long-baseline interferometry, accretion disk spectroscopy, star dynamics or early inspiral gravitational waves --- these objects represent phenomenologically viable ``mimickers'' of Schwarzschild black holes, similar but not identical to their singular counterparts. 
The search for signatures of these subtle deviations could then mark the dawn of a new channel for quantum gravity phenomenology.

\section*{Acknowledgements}

We wish to thank Edgrado Franzin, Enrico Barausse and Mario Herrero-Valea for the precious discussions and their comments on an early draft of the manuscript.
The authors acknowledge funding from the Italian Ministry of Education and Scientific Research (MIUR) under the grant PRIN MIUR 2017-MB8AEZ.

\appendix

\section{Optical scalars} \label{app:scalars}

A coordinate-independent way to characterise the \aet~congruence is through the optical scalars: 
\be
\text{expansion} \quad \theta &= \nabla_a u^a\, , \\
\text{shear squared} \quad \sigma^2 \qq{with} 
 \sigma_{ab}&= \nabla_{(a} u_{b)} - u_{(a} a_{b)} - \frac{\theta}{3}P_{ab}\, ,\\
\text{twist squared} \quad \omega^2 \qq{with} \omega_{ab}&= \nabla_{[b} u_{c]} - u_{[a} a_{b]}  \, .
\ee
Other interesting scalars are $ u_a \, \chi^a$ and $a_a \, \chi^a$, as they are associated with properties of the UHs.

Since the \aet~is hypersurface-orthogonal, Frobenius' theorem implies that the twist vanishes.
(Note that $\omega^2 \propto (u_{[a}\nabla_{b} u_{c]})^2$.)
Moreover
\be
a_a \, \chi^a = \sqrt{-a^2} = y \left( u_a \, \chi^a \right)'\, .
\ee

When evaluated on the Ansatz of \cref{eq:met,eq:aet}, one finds
\be
\theta &= y' + 2y \frac{R'}{R} 
\ee
and a similar, though lengthier, expression for $\sigma^2$.
All this quantities thus depend algebraically on the functions $F(r),\ R(r),\ A(r)$ and their first derivatives, in a way that renders the following statement manifestly true: when $F(r),\ R(r),\ A(r)$ are of class $\mathcal{C}^1$ and bounded, all the scalars introduced above are $\mathcal{C}^1$ and bounded.
We have computed them explicitly for the singular solution and found that they are ill-behaved at the origin; and then on the connected and on the non-connected non-singular configurations, checking that they are indeed well-behaved everywhere --- in particular, at the origin, at the UHs and at the KHs.

\section{2D expansions}\label{app:expansions}

In order to make contact with the arguments of \cite{carballo-rubio_geodesically_2022}, we complement our analysis with a discussion on the local characterisation of horizons.

We start by considering a closed, spacelike 2-surface $\mathscr{S}^2$. 
The subspace of the tangent space that is orthogonal to the tangent space of $\mathscr{S}^2$ is spanned by two vectors that can be taken timelike, future-pointing and spacelike, outward-pointing --- respectively.
In our case, a simple choice for $\mathscr{S}^2$ is any sphere centred at the origin. The two vectors are then the \aet~and the vector $s_a$ of \cref{eq:s} used to define the tangential pressure.

The induced metric on $\mathscr{S}^2$ is
\be
h_{ab} = g_{ab} - u_a u_b + s_a s_b
\ee
and can be used to define the scalars
\be
\theta^{(X)} = h^{ab} \nabla_a X_b \qq{with} X = \{u, s\}\, .
\ee
These are expansions, but should not be confused with the optical scalar $\theta$, which is defined in terms of a three-dimensional transverse metric. 
$\theta^{(u)}$ and $\theta^{(s)}$, and in particular their signs, determine whether $\mathscr{S}^2$ is a universal (marginally) trapped surface.

With our Ans\"{a}tze of \cref{eq:met,eq:aet}, we have
\be
\theta^{(u)} = 2y \frac{R'}{R} \qq{and} \theta^{(s)} = 2 ( u_a \, \chi^a) \frac{R'}{R}\, .
\ee

Recall that $y<0$.
Hence, on the singular solution \cref{eq:exact_met,eq:exact_A}, $\theta^{(u)}$ is always negative, i.e.~the future-directed congruence is always converging, while $\theta^{(s)}$ has the sign of $u_a \, \chi^a$.
Thus, $u_a \, \chi^a = 0$ marks a universal trapping horizon.
Note that both expansions diverge as $r \to  0$, meaning that $r=0$ is a caustic. Penrose's theorem then implies that this is in fact a singularity, in the sense that the spacetime is not geodesically complete.

On the simply connected configurations of \cref{eq:regConm,eq:regConae} we still have that $\theta^{(u)}<0$ and that $\theta^{(s)}$ has the sign of $u_a \, \chi^a$, but we know that in this case $u_a \, \chi^a = 0$ has multiple roots and in particular it is positive in a neighbourhood of $r=0$. Hence there exist multiple universal trapping horizons.
Further note that in this case $r=0$ is not a caustic anymore, since $\theta^{(u)} \to 0$ and $\theta^{(s)} \to 1$ as $r\to 0$.

In the non-connected configuration the sign of the two expansions depends also on $R'/R$, which is positive in our universe but negative in the other. I.e.~both congruences vanish and change sign at the wormhole mouth $r=0$.


\bibliographystyle{JHEP.bst}
\bibliography{Horava-Lifshitz}

\end{document}